\documentclass[byrevtex,groupedaddress,prb,showpacs,oneside,final,preprint,titlepage,11pt,onecolumn]{revtex4}
\usepackage{amsfonts}
\usepackage{amsmath}
\usepackage{amssymb}
\usepackage{graphicx}
\usepackage{dcolumn}
\usepackage{bm}

\input{tcilatex}

\begin{document}

\preprint{}
\title{Decoherence of a driven multilevel quantum system interacting with a
multi-bath reservoir}
\author{Zhongyuan Zhou$^{1,2}$}
\author{Shih-I Chu$^{1}$}
\author{Siyuan Han$^{2}$}
\affiliation{$^{1}$Department of Chemistry, University of Kansas, Lawrence, KS 66045\\
$^{2}$Department of Physics and Astronomy, University of Kansas, Lawrence,
KS 66045}

\begin{abstract}
A general theory is presented for the treatment of decoherence of a
multilevel quantum system (with many degrees of freedom) interacting with
multi-bath reservoir and driven by ac fields. In this approach, the system
is described by a reduced density operator and the multi-bath reservoir is
characterized by a number of spectral densities. The reduced density
operator is governed by the master equation in which the effect of ac
driving fields and the leakage to non-computational states are included. The
theory is applied to the study of decoherence of a two-dimensional (2D)
SQUID flux qubit coupled to control and readout circuits. The predicted
results are in very good agreement with available experimental results in
the absence of driving fields and with the analytic results of a dissipative
two-level system in the presence of weak driving fields. The relaxation and
decoherence times versus the parameters and temperature of the control and
readout circuits are also explored in details to facilitate the optimization
of the 2D SQUID qubit.
\end{abstract}

\received[Received: ]{\today }
\maketitle

\section{Introduction}

In the past few years, solid-state qubits based on superconducting devices
are of particular interest for quantum computation because of their
advantages of large-scale integration, flexibility in design, and easy
connection to conventional electronic circuits \cite{Mooij1999}. Significant
progress has been made on physical implementation of quantum computation
based on superconducting qubits. Quantum coherence has been successfully
demonstrated in a variety of superconducting single-qubit systems \cite%
{Nakamura1999,Vion2002,Yu2002,Martinis2002,Friedman2000,Wal2000,Chiorescu2003,Chiorescu2004}
and coupled two-qubit systems. \cite%
{Pashkin2003,Yamamoto2003,Berkley2003,McDermott05} However, all the
superconducting qubits demonstrated in the experiments so far have
relatively short coherence time and high probability of gate errors. \cite%
{Nakamura1999,Vion2002,Yu2002,Martinis2002,Chiorescu2003,Pashkin2003,Yamamoto2003,Chiorescu2004}
One of the causes of these problems is the intrinsic gate error resulting
from leakage to non-computational states due to the typical multilevel
structures of superconducting qubits. \cite{Fazio99,Zhou2002}\textit{\ }This
kind of gate error has been explored \cite{Fazio99,Zhou2002,zhou2005PRL-ITA}
and can be minimized by using appropriate working parameters for the qubits
with their device parameters given \textit{in prior}. \cite{zhou2005PRL-ITA}
Another cause is the extrinsic gate error arising from coupling between the
qubits and environment resulting in decoherence such as relaxation and
dephasing. \cite{Makhlin2001,Zhou2004} Due to the unavoidable coupling with
environment, the superconducting qubits always suffer from such kind of
extrinsic gate error. Thus the environment-induced decoherence is one of the
main obstacles to the practical application of superconducting qubits in
quantum computation. \cite%
{Vion2002,Astafiev04,Ithier05,Makhlin2001,Mooij1999}

The typical environment in superconducting qubits is electronic circuits
used for control and readout of the qubits. Although the decoherence of
superconducting qubits induced by such kind of environment has been
extensively investigated both theoretically\emph{\ }\cite%
{Burkard2004,Mooij1999,Makhlin2001,Burkard05-1,Goorden04,Makhlin04,Orlando2002,Harlingen04,Zhou2004,Shresta05,Anastopoulos2000,Robertson05,Cheng2004,Storcz2003,Governale2001,Tian2002,Xu2005,Falci05}
and experimentally.\emph{\ }\cite%
{Chiorescu2004,Astafiev04,Nakamura02,Lehnert03,Duty04,Lishaoxiong06,Bertet06,Vion2002,Berkley03-1,Dutta2004,Harlingen04,Robertson05}%
, almost all the investigations so far are for the qubits in the absence of
driving fields (free decay). However, in superconducting-qubit based quantum
computation, ac fields (e.g., microwave fields) are usually used to
manipulate the qubit's states. \cite%
{Nakamura1999,Vion2002,Yu2002,Martinis2002,Friedman2000,Wal2000,Chiorescu2003,Pashkin2003,Yamamoto2003,Berkley2003,Chiorescu2004,McDermott05,Goorden04}
Due to the coupling of the qubits with the fields the effect of driving
fields and leakage to non-computational states may be quite large depending
on the field strength \cite{Fazio99,Zhou2002,zhou2005PRL-ITA} and thus the
dynamics of driven qubits may be quite different from that of qubits in free
decay. Recent experiment \cite{Ithier05} shows that the decoherence time of
a superconducting qubit is significantly increased in the presence of a
resonantly ac driving field. Thus a through investigation of decoherence of
realistic superconducting qubits needs to include effect of driving fields
and leakage to non-computational states. \cite{Goorden04}

Furthermore, building a practical quantum computer requires simultaneous
operations of a large number of multiqubit gates in a coupled multiqubit
system. \cite{Barenco1995,Nielsen2000,zhou2006} On one hand, the coupled
multiqubit system may have many degrees of freedom, on the other hands, due
to complexity and diversity of the superconducting circuits, the environment
may be a multibath one. In this case, the coupled multiqubit system may
interact with the environment through all the degrees of freedom \emph{%
simultaneously}, resulting in significantly different decoherence from that
interacting with the environment through one degree of freedom of single
qubits. Therefore, to investigate the decoherence of coupled superconducting
qubits in realistic gate operations of quantum computation, one needs an
approach for a quantum system of many degrees of freedom interacting with a
multibath reservoir and driven by ac fields.

In this paper, we propose a general theory for the treatment of decoheerence
of a multilevel quantum system of many degrees of freedom interacting with a
multibath reservoir and driven by ac fields. In this theory, the multibath
reservoir are characterized by a number of spectral densities. \cite%
{Leggett87,Weiss1999,Devoret97} For superconducting qubits coupled to
electronic circuits, the spectral densities at finite temperature can be
calculated in terms of recently proposed approaches \cite%
{Burkard2004,Devoret97} together with the quantum fluctuation-dissipation
theorem. \cite{Weiss1999,Devoret97} The system is described by a reduced
density operator. The reduced density operator is governed by the master
equation in which the effect of driving fields and leakage due to the
driving field and reservoir are included. This theory is used to simulate
the dynamic process of a two-dimensional (2D) superconducting quantum
interface device (SQUID) flux qubit coupled to control and readout circuits
in the absence of driving fields (free decay). The results are in very good
agreement with the available experimental results. It is also applied to
investigate the effect of driving fields on the decoherence of the 2D SQUID
qubit coupled to the control and readout circuits and driven by a resonant
microwave field. The results agree well with the analytical results of the
dissipative two-level system in the case of weak driving fields. To optimize
the 2D SQUID qubit, the changes of relaxation and decoherence times versus
the parameters and temperature of the control and readout circuits are
explored.

\section{Master equation of a driven open quantum system}

\subsection{General form of master equation for a driven open quantum system}

In general, an open quantum system is described by a generalized master
equation of non-Markovian process. \cite%
{Brinati94,Smirnov2003,Hartmann2000,Shresta05} However, in the case of weak
damping and weak driving fields, which is the case for most of
superconducting (charge, flux, and phase) qubits, the generalized
non-Markovian master equation is equivalent to the Markovian master
equation. \cite{Brinati94,Hartmann2000} Thus we will present the dissipative
theory for Markovian process.

Let us consider a global system consisting of a quantum system surrounded by
a reservoir and driven by an ac field. If the interaction of the driving
field and reservoir is neglected, the Hamiltonian of the global system can
be written as%
\begin{equation}
H_{G}=H_{S}+H_{R}+H_{I}+H_{F},  \label{a1}
\end{equation}%
where, $H_{S}$ is the Hamiltonian of the quantum system, $H_{R}$ is the
Hamiltonian of the reservoir, $H_{I}$ is the interaction of the quantum
system and reservoir, and $H_{F}$ is the interaction of the quantum system
and driving field. Obviously $H_{R}$ commutes with both $H_{S}$ and $H_{F}$.
For the case of weak damping and weak driving field considered here, the
interactions $H_{I}$ and $H_{F}$ are proportional to the system's coordinate
operators linearly and thus they also commute with each other. \cite%
{Louisell1973}

In Schr\"{o}dinger picture, the motion of the global system is described by 
\textit{Liouville-von Neumann} equation \cite{Breuer2002}%
\begin{equation}
\frac{d\eta (t)}{dt}=-\frac{i}{\hbar }\left[ H_{G}(t),\eta (t)\right] =-i%
\mathcal{L}(t)\eta (t),  \label{a8}
\end{equation}%
where, $\eta (t)$ is the density operator of the global system and $\mathcal{%
L}(t)$ is the Liouville superoperator corresponding to the Hamiltonian of
the global system defined by%
\begin{equation}
\mathcal{L=}\frac{1}{\hbar }\left[ H_{G},\right] =\mathcal{L}_{S}+\mathcal{L}%
_{R}+\mathcal{L}_{I}+\mathcal{L}_{F}.  \label{a9}
\end{equation}%
Here, $\mathcal{L}_{q}=\left[ H_{q},\right] /\hbar $ is the Liouville
superoperator corresponding to the Hamiltonian $H_{q}$ for $q=S$, $R$, $I$,
and $F$. The density operator of the global system satisfies Tr$_{S,R}\eta
(t)=1$, where Tr$_{S,R}$ is the trace over both the quantum system and
reservoir.

In the global system, the quantum system performs a dissipative process due
to the coupling with reservoir. This process can be characterized by a
reduced density operator $\rho \left( t\right) $. It is calculated by
tracing $\eta (t)$ over the reservoir%
\begin{equation}
\rho (t)=\text{Tr}_{R}\eta (t).  \label{a17-1}
\end{equation}%
In the case of weak field and weak damping, $\rho (t)$ is governed by the
master equation which, in Schr\"{o}dinger picture, is given by (see Appendix
A for details) 
\begin{equation}
\frac{d\rho (t)}{dt}=-i\left[ \mathcal{L}_{S}+\mathcal{L}_{F}(t)\right] \rho
(t)+\mathcal{D}_{I}(t)\rho (t),  \label{a25}
\end{equation}%
where, $\mathcal{D}_{I}\left( t\right) $ is the dissipation superoperator
given by Eq. (\ref{a23}). On the right-hand side of Eq. (\ref{a25}), the
first term describes the coherent dynamics of the unperturbed quantum
system, the second term represents the pumping dynamics of the driving
field, and the third term which is called the dissipator \cite{Breuer2002}
characterizes the dissipative dynamics due to the interaction of the quantum
system and reservoir. The dissipator includes all the information of
dissipative dynamics. \cite{Toutounji05} Note that Eq. (\ref{a25}) is
determined only by the operators at present and no longer depends on the
operators in the past. Thus it describes a Markovian process of the quantum
system \cite{Louisell1973}. The Markovian process is irreversible because
the energy transferred to the reservoir can no longer return to the quantum
system completely. In the case of weak damping and weak driving fields the
Markovian approximation is equivalent to the non-Markovian approach. \cite%
{Brinati94,Hartmann2000}

\subsection{Master equation of a driven quantum system of many degrees of
freedom interacting with a multibath reservoir}

Let us consider a quantum system of many degrees of freedom encompassed by a
multibath reservoir and driven by an ac field. The multibath can be
separated into a number of independent baths. They have different properties
and interact with the system via different degrees of freedom. If
interactions between the baths are incorporated into the Hamiltonian of the
reservoir, the total interaction $H_{I}$ of the system and the multibath
reservoir is given by \cite{Louisell1973,Gaspard1999}%
\begin{equation}
H_{I}(t)=\sum_{\mu }X_{\mu }(t)\varsigma _{\mu }(t),  \label{m20}
\end{equation}%
where, $\varsigma _{\mu }(t)$ is the operator of the $\mu $th bath of the
reservoir, $X_{\mu }(t)$ is the system operator corresponding to the $\mu $%
th bath, and the sum to $\mu $ is over all the baths of the reservoir and
all the degrees of freedom of the system. In general, the interaction $%
H_{I}(t)$ is a Hermitian operator but $X_{\mu }$ and $\varsigma _{\mu }$ may
be non-Hermitian. In the case of weak damping, $X_{\mu }(t)$ is reduced to
the system coordinate operator and $\varsigma _{\mu }(t)$ is reduced to the
fluctuation force of the $\mu $th bath. \cite{Leggett87,Weiss1999} For a
driven quantum system of many degrees of freedom interacting with a
multibath reservoir, the reduced density operator of the system still
satisfies the maser equation given by Eq. (\ref{a25}). Using Eqs. (\ref{a9}%
), (\ref{m20}), (\ref{a14}), (\ref{a23}), and (\ref{a23-2}), the dissipator $%
\mathcal{D}_{I}(t)\rho (t)$ in Eq. (\ref{a25}) is now given by%
\begin{eqnarray}
\mathcal{D}_{I}(t)\rho (t) &=&\frac{1}{\hbar ^{2}}\sum_{\mu ,\nu
}\int_{0}^{t}\left[ X_{\mu }\left( t\right) \rho \left( t\right) \widetilde{X%
}_{\nu }\left( \tau \right) \mathcal{J}_{\mu \nu }\left( \tau \right) \right.
\notag \\
&&-X_{\mu }\left( t\right) \widetilde{X}_{\nu }\left( \tau \right) \rho
\left( t\right) \mathcal{J}_{\nu \mu }^{\dagger }\left( -\tau \right)  \notag
\\
&&-\rho \left( t\right) \widetilde{X}_{\nu }\left( \tau \right) X_{\mu
}\left( t\right) \mathcal{J}_{\mu \nu }\left( \tau \right)  \notag \\
&&\left. +\widetilde{X}_{\nu }\left( \tau \right) \rho \left( t\right)
X_{\mu }\left( t\right) \mathcal{J}_{\nu \mu }^{\dagger }\left( -\tau
\right) \right] d\tau ,  \label{m26-2-1}
\end{eqnarray}%
where, $\widetilde{X}_{\mu }$ is the representation of $X_{\mu }$ in the
interaction picture given by%
\begin{equation}
\widetilde{X}_{\mu }\left( -\tau \right) =\exp \left( \frac{i}{\hbar }%
H_{S}\tau \right) X_{\mu }\left( t+\tau \right) \exp \left( -\frac{i}{\hbar }%
H_{S}\tau \right) ,  \label{m24}
\end{equation}%
and $\mathcal{J}_{\mu \nu }\left( \tau \right) $ and $\mathcal{J}_{\nu \mu
}^{\dagger }\left( -\tau \right) $ are the two kinds of correlation
functions of the fluctuation forces between the $\mu $th and $\nu $th baths.
The correlation function $\mathcal{J}_{\mu \nu }\left( \tau \right) $ is
defined by%
\begin{eqnarray}
\mathcal{J}_{\mu \nu }\left( \tau \right) &=&\text{Tr}_{R}\left[ \widetilde{%
\varsigma }_{\mu }(t)\sigma \left( R\right) \widetilde{\varsigma }_{\nu
}(t-\tau )\right]  \notag \\
&=&\left\langle \widetilde{\varsigma }_{\mu }(t)\widetilde{\varsigma }_{\nu
}(t-\tau )\right\rangle ,  \label{m25}
\end{eqnarray}%
where, $\widetilde{\varsigma }_{\mu }$ is the representation of $\varsigma
_{\mu }$ in the interaction picture defined by%
\begin{equation}
\widetilde{\varsigma }_{\mu }\left( \tau \right) =\exp \left( \frac{i}{\hbar 
}H_{R}\tau \right) \varsigma _{\mu }\left( \tau \right) \exp \left( -\frac{i%
}{\hbar }H_{R}\tau \right) .  \label{m24-1}
\end{equation}%
The correlation function $\mathcal{J}_{\nu \mu }^{\dagger }\left( -\tau
\right) $ is the complex transpose of $\mathcal{J}_{\mu \nu }\left( \tau
\right) $. In general, since $\widetilde{\varsigma }_{\mu }(t)$ and $%
\widetilde{\varsigma }_{\nu }(t-\tau )$ may not be Hermitian operators
and/or they do not commute with each other, $\mathcal{J}_{\mu \nu }\left(
\tau \right) $ is a non-Hermitian matrix and $\mathcal{J}_{\nu \mu
}^{\dagger }\left( -\tau \right) \neq \mathcal{J}_{\mu \nu }\left( \tau
\right) $.

If the $\mu $th bath does not interact with the $\nu $th bath, the
correlation between the two baths is zero. In this case, the correlation
function $\mathcal{J}_{\mu \nu }\left( \tau \right) $ can be written as%
\begin{equation}
\mathcal{J}_{\mu \nu }\left( \tau \right) =\mathcal{J}_{\mu \mu }\left( \tau
\right) \delta _{\mu \nu },  \label{m26}
\end{equation}%
where, $\mathcal{J}_{\mu \mu }\left( \tau \right) $ is the autocorrelation
function. The autocorrelation function $\mathcal{J}_{\mu \mu }\left( \tau
\right) $ represents the correlation of the bath state at time $t-\tau $ to
the bath state at time $t$, while the autocorrelation function $\mathcal{J}%
_{\mu \mu }^{\dagger }\left( -\tau \right) $ represents the correlation of
the bath state at time $t$ to the bath state at time $t-\tau $. In general, $%
\mathcal{J}_{\mu \mu }^{\dagger }\left( -\tau \right) \neq \mathcal{J}_{\mu
\mu }\left( \tau \right) $, which signifies the irreversibility in time of
the correlation function. From Eq. (\ref{m25}), the irreversibility is
attributed to the non-Hermitian property and/or noncommutable property of
the reservoir operators.

From Eq. (\ref{m26-2-1}), the effect of reservoir on the system is
represented by the correlation function $\mathcal{J}(t)$. It is difficult to
calculate the correlation function directly from the fluctuation forces of
the reservoir. In reality, the effect of reservoir on the system can be
equivalently characterized by a real and measurable spectral density of the
reservoir \cite{Leggett87,Weiss1999,Devoret97}. The relation between the
correlation function $\mathcal{J}(t)$ and the spectral density $J_{\varsigma
}\left( \omega \right) $ is given by the Wiener-Khintchine theorem \cite%
{Kubo91}%
\begin{equation}
\mathcal{J}(t)=\frac{1}{2\pi }\int_{-\infty }^{+\infty }J_{\varsigma }\left(
\omega \right) \exp \left( i\omega t\right) d\omega ,  \label{m26-3}
\end{equation}%
and%
\begin{equation}
J_{\varsigma }\left( \omega \right) =\int_{-\infty }^{+\infty }\mathcal{J}%
(t)\exp \left( -i\omega t\right) dt.  \label{m26-4}
\end{equation}%
For solid-state qubits, particularly for the superconducting qubits
consisting of electronic circuits, the spectral density at zero temperature
can be calculated in terms of the recently proposed approaches. \cite%
{Burkard2004,Devoret97} The spectral density at finite temperature can be
obtained from the zero-temperature spectral density by means of the quantum
fluctuation-dissipation theorem \cite{Weiss1999,Devoret97}.

\section{Representation of master equation in Hilbert space}

\subsection{Representation in an arbitrary Hilbert space}

Suppose $\left\{ \left\vert n\right\rangle \right\} $ is a complete basis
set of the system. In the Hilbert space spanned by $\left\{ \left\vert
n\right\rangle \right\} $, the reduced density operator $\rho $ is
represented by a density matrix with matrix elements given by%
\begin{equation}
\rho _{mn}\left( t\right) =\left\langle m|\rho \left( t\right)
|n\right\rangle .  \label{n1}
\end{equation}%
The diagonal matrix element $\rho _{mm}$ and off-diagonal matrix element $%
\rho _{mn}$ $(m\neq n)$ are the population of state $\left\vert
m\right\rangle $ and coherence of states $\left\vert m\right\rangle $ and $%
\left\vert n\right\rangle $, respectively. From Eqs. (\ref{a25}) and (\ref%
{m26-2-1}), the density matrix element $\rho _{mn}\left( t\right) $
satisfies the master equation%
\begin{eqnarray}
\frac{d\rho _{mn}\left( t\right) }{dt} &=&-i\sum_{m^{\prime }n^{\prime }}%
\left[ \mathcal{L}_{mn,m^{\prime }n^{\prime }}^{S}+\mathcal{L}_{mn,m^{\prime
}n^{\prime }}^{F}\right] \rho _{m^{\prime }n^{\prime }}  \notag \\
&&+\sum_{m^{\prime }n^{\prime }}\mathcal{D}_{mn,m^{\prime }n^{\prime
}}^{I}\rho _{m^{\prime }n^{\prime }},  \label{n3}
\end{eqnarray}%
where, $\mathcal{L}_{mn,m^{\prime }n^{\prime }}^{S}$\ and $\mathcal{L}%
_{mn,m^{\prime }n^{\prime }}^{F}$ are the matrix elements of $\mathcal{L}%
_{S} $ and $\mathcal{L}_{F}$, respectively, and $\mathcal{D}_{mn,m^{\prime
}n^{\prime }}^{I}$ is the dissipation matrix element of the dissipation
superoperator $\mathcal{D}_{I}$ describing the effect of the reservoir on
the system. They are given by 
\begin{equation}
\mathcal{L}_{mn,m^{\prime }n^{\prime }}^{\Theta }=\frac{1}{\hbar }\left[
H_{mm^{\prime }}^{\Theta }\delta _{n^{\prime }n}-H_{n^{\prime }n}^{\Theta
}\delta _{mm^{\prime }}\right] ,  \label{n4}
\end{equation}%
and%
\begin{eqnarray}
\mathcal{D}_{mn,m^{\prime }n^{\prime }}^{I} &=&\frac{1}{\hbar ^{2}}\sum_{\mu
,\nu }\int_{0}^{t}\left[ X_{mm^{\prime }}^{\mu }\left( t\right) \widetilde{X}%
_{n^{\prime }n}^{\nu }\left( \tau \right) \mathcal{J}_{\mu \nu }\left( \tau
\right) \right.  \notag \\
&&-\delta _{nn^{\prime }}\sum_{k}X_{mk}^{\mu }\left( t\right) \widetilde{X}%
_{km^{\prime }}^{\nu }\left( \tau \right) \mathcal{J}_{\nu \mu }^{\dagger
}\left( -\tau \right)  \notag \\
&&-\delta _{mm^{\prime }}\sum_{k}\widetilde{X}_{n^{\prime }k}^{\nu }\left(
\tau \right) X_{kn}^{\mu }\left( t\right) \mathcal{J}_{\mu \nu }\left( \tau
\right)  \notag \\
&&+\left. \widetilde{X}_{mm^{\prime }}^{\nu }\left( \tau \right)
X_{n^{\prime }n}^{\mu }\left( t\right) \mathcal{J}_{\nu \mu }^{\dagger
}\left( -\tau \right) \right] d\tau ,  \label{n8}
\end{eqnarray}%
where, $H_{mn}^{\Theta }=\left\langle m\left\vert H_{\Theta }\right\vert
n\right\rangle $ for $\Theta =S$ and $F$, and $X_{mn}^{\alpha }\left(
t\right) =\left\langle m\left\vert X_{\alpha }\left( t\right) \right\vert
n\right\rangle $ and $\widetilde{X}_{mn}^{\alpha }\left( t\right)
=\left\langle m\left\vert \widetilde{X}_{\alpha }\left( t\right) \right\vert
n\right\rangle $ for $\alpha =\mu $ and $\nu $.

\subsection{Representation in the Hilbert space of Hamiltonian eigenstates}

Suppose $E_{n}$ and $\left\vert n\right\rangle $ are respectively the
eigenvalue and eigenfunction of $H_{S}$ obtained by solving the eigenvalue
equation $H_{S}\left\vert n\right\rangle =E_{n}\left\vert n\right\rangle $.
In the Hilbert space spanned by the eigenstates $\left\{ \left\vert
n\right\rangle \right\} $, the density matrix element and the master
equation are still given by Eq. (\ref{n1}) and Eq. (\ref{n3}), respectively.
But now the matrix elements $\mathcal{L}_{mn,m^{\prime }n^{\prime }}^{S}$
and $\mathcal{L}_{mn,m^{\prime }n^{\prime }}^{F}$ are simplified to%
\begin{equation}
\mathcal{L}_{mn,m^{\prime }n^{\prime }}^{S}=\omega _{mn}\delta _{mm^{\prime
}}\delta _{nn^{\prime }},  \label{m31}
\end{equation}%
and%
\begin{equation}
\mathcal{L}_{mn,m^{\prime }n^{\prime }}^{F}=\frac{1}{\hbar }\left[
H_{mm^{\prime }}^{F}(t)\delta _{nn^{\prime }}-H_{n^{\prime }n}^{F}(t)\delta
_{mm^{\prime }}\right] ,  \label{m32}
\end{equation}%
where, $\omega _{mn}=\left( E_{m}-E_{n}\right) /\hbar $ and $%
H_{mn}^{F}(t)=\left\langle m\left\vert H_{F}(t)\right\vert n\right\rangle $.
The matrix element $\mathcal{D}_{mn,m^{\prime }n^{\prime }}^{I}$ in Eq. (\ref%
{n8}) is also simplified to%
\begin{eqnarray}
\mathcal{D}_{mn,m^{\prime }n^{\prime }}^{I} &=&\frac{1}{\hbar ^{2}}\sum_{\mu
,\nu }\int_{0}^{t}\left[ X_{mm^{\prime }}^{\mu }\left( t\right) X_{n^{\prime
}n}^{\nu }\left( t^{\prime }\right) e^{i\omega _{nn^{\prime }}\tau }\mathcal{%
J}_{\mu \nu }\left( \tau \right) \right.  \notag \\
&&-\delta _{nn^{\prime }}\sum_{k}X_{mk}^{\mu }\left( t\right) X_{km^{\prime
}}^{\nu }\left( t^{\prime }\right) e^{i\omega _{m^{\prime }k}\tau }\mathcal{J%
}_{\nu \mu }^{\dagger }\left( -\tau \right)  \notag \\
&&-\delta _{mm^{\prime }}\sum_{k}X_{n^{\prime }k}^{\nu }\left( t^{\prime
}\right) X_{kn}^{\mu }\left( t\right) e^{i\omega _{kn^{\prime }}\tau }%
\mathcal{J}_{\mu \nu }\left( \tau \right)  \notag \\
&&+\left. X_{mm^{\prime }}^{\nu }\left( t^{\prime }\right) X_{n^{\prime
}n}^{\mu }\left( t\right) e^{i\omega _{m^{\prime }m}\tau }\mathcal{J}_{\nu
\mu }^{\dagger }\left( -\tau \right) \right] d\tau ,  \label{m39}
\end{eqnarray}%
where, $t^{\prime }=t-\tau $.

If the Hamiltonian eigenfunctions $\left\{ \left\vert n\right\rangle
\right\} $ are real, one has $H_{mn}^{F}=H_{nm}^{F}$ for the Hermitian
operator $H_{F}(t)$. From Eqs. (\ref{m31}) and (\ref{m32}), the matrix
elements $\mathcal{L}_{mn,m^{\prime }n^{\prime }}^{\Theta }$ for $\Theta =S$
and $F$ have the following symmetric relations%
\begin{equation}
\mathcal{L}_{nm,n^{\prime }m^{\prime }}^{\Theta }=-\mathcal{L}_{mn,m^{\prime
}n^{\prime }}^{\Theta },  \label{m42-1}
\end{equation}%
and%
\begin{equation}
\mathcal{L}_{m^{\prime }n^{\prime },mn}^{\Theta }=\mathcal{L}_{mn,m^{\prime
}n^{\prime }}^{\Theta }.  \label{m42-2}
\end{equation}%
From Eq. (\ref{m42-1}) one has $\mathcal{L}_{mm,nn}^{\Theta }=0.$

If $X_{\mu }$\ is time-independent, the matrix elements $\mathcal{D}%
_{mn,m^{\prime }n^{\prime }}^{I}$ can be further simplified. Introduce a
function $F_{\mu \nu }(\omega ,t)$ by%
\begin{equation}
F_{\mu \nu }(\omega ,t)=\int_{0}^{t}e^{i\omega \tau }\mathcal{J}_{\mu \nu
}\left( \tau \right) d\tau .  \label{a51}
\end{equation}%
Applying Eq. (\ref{a51}) to Eq. (\ref{m39}) one obtains%
\begin{eqnarray}
\mathcal{D}_{mn,m^{\prime }n^{\prime }}^{I} &=&\frac{1}{\hbar ^{2}}\sum_{\mu
,\nu }\left[ X_{mm^{\prime }}^{\mu }X_{n^{\prime }n}^{\nu }F_{\mu \nu
}(\omega _{nn^{\prime }},t)\right.  \notag \\
&&-\delta _{nn^{\prime }}\sum_{k}X_{mk}^{\mu }X_{km^{\prime }}^{\nu }F_{\mu
\nu }^{\dag }(\omega _{km^{\prime }},t)  \notag \\
&&-\delta _{mm^{\prime }}\sum_{k}X_{n^{\prime }k}^{\nu }X_{kn}^{\mu }F_{\mu
\nu }(\omega _{kn^{\prime }},t)  \notag \\
&&+\left. X_{mm^{\prime }}^{\nu }X_{n^{\prime }n}^{\mu }F_{\mu \nu }^{\dag
}(\omega _{mm^{\prime }},t)\right] ,  \label{a52}
\end{eqnarray}%
where $F_{\mu \nu }^{\dag }(\omega ,t)$ is the conjugate transpose of $%
F_{\mu \nu }(\omega ,t)$. If the interaction between any two baths is
negligible, $\mathcal{D}_{mn,m^{\prime }n^{\prime }}^{I}$ can be further
simplified by using Eqs. (\ref{m26}) and (\ref{a51}). In this case, the
correlation terms between different baths in Eq. (\ref{a52}) vanishes and
thus the dissipation matrix element of the multibath reservoir is a sum of
individual dissipation matrix element of each bath of the reservoir.

IF time-dependent part of $X_{\mu }$ is separable from system operators, $%
X_{\mu }$ can be expressed in a general form of Fourier series as%
\begin{equation}
X_{\mu }\left( t\right) =\sum_{\lambda }\chi _{\lambda }^{\mu }\exp \left(
i\omega _{\lambda }^{\mu }t\right) ,  \label{a61}
\end{equation}%
where, $\chi _{\lambda }^{\mu }$ is a time-independent operator that only
depends on system operators.

Substituting Eq. (\ref{a61}) into Eq. (\ref{m39}) and using Eq. (\ref{a51}),
one has%
\begin{eqnarray}
\mathcal{D}_{mn,m^{\prime }n^{\prime }}^{I} &=&\frac{1}{\hbar ^{2}}\sum_{\mu
,\nu }\sum_{\lambda _{1}\lambda _{2}}\exp \left[ i\left( \omega _{\lambda
_{1}}^{\mu }+\omega _{\lambda _{2}}^{\nu }\right) t\right]  \notag \\
&&\times \left[ \chi _{\lambda _{1},mm^{\prime }}^{\mu }\chi _{\lambda
_{2},n^{\prime }n}^{\nu }F_{\mu \nu }\left( \omega _{\lambda _{2},nn^{\prime
}}^{\nu -},t\right) \right.  \notag \\
&&-\delta _{nn^{\prime }}\sum_{k}\chi _{\lambda _{1},mk}^{\mu }\chi
_{\lambda _{2},km^{\prime }}^{\nu }F_{\mu \nu }^{\dag }\left( \omega
_{\lambda _{2},km^{\prime }}^{\nu +},t\right)  \notag \\
&&-\delta _{mm^{\prime }}\sum_{k}\chi _{\lambda _{2},n^{\prime }k}^{\nu
}\chi _{\lambda _{1},kn}^{\mu }F_{\mu \nu }\left( \omega _{\lambda
_{2},kn^{\prime }}^{\nu -},t\right)  \notag \\
&&+\left. \chi _{\lambda _{2},mm^{\prime }}^{\nu }\chi _{\lambda
_{1},n^{\prime }n}^{\mu }F_{\mu \nu }^{\dag }\left( \omega _{\lambda
_{2},mm^{\prime }}^{\nu +},t\right) \right] ,  \label{a62}
\end{eqnarray}%
where, $\omega _{\lambda _{2},nn^{\prime }}^{\nu \pm }=\omega _{nn^{\prime
}}\pm \omega _{\lambda _{2}}^{\nu }$ and $\chi _{\lambda ,mn}^{\alpha
}=\left\langle m\left\vert \chi _{\lambda }^{\alpha }\right\vert
n\right\rangle $ is the matrix element of the operator $\chi _{\lambda
}^{\alpha }$.

Furthermore, if $\chi _{\lambda }^{\alpha }$ is a time-independent operator
that does not depend on system operators we have $\chi _{\lambda
,mn}^{\alpha }=\left\langle m\left\vert \chi _{\lambda }^{\alpha
}\right\vert n\right\rangle =\chi _{\lambda }^{\alpha }\delta _{mn}$. In
this case, from Eq. (\ref{a62}) one has $\mathcal{D}_{mn,m^{\prime
}n^{\prime }}^{I}=0$. Thus if the interaction of the system and reservoir
does not depend on system operators, the reservoir does not have any effect
on the system and thus they are decoupled.

\section{Dissipation of a driven quantum system due to a thermal bath}

One of the most popular and important reservoirs is a thermal bath. Due to
the interaction with the thermal bath, the quantum system transits from one
thermodynamic equilibrium state to another equilibrium state. \cite%
{Orlando2002} During this process, the spontaneous decay and stimulated
transition follow the detailed balance principle. As has been demonstrated,
if the interactions between the independent baths are neglected the
dissipation matrix element of the multibath reservoir is a sum of individual
dissipation matrix element of each bath. Thus in this section we will
present the dissipative theory for a driven quantum system interacting with
a thermal bath.

\subsection{Lamb shift matrix and damping rate matrix}

Suppose the driven quantum system is surrounded only by a sufficiently large
thermal bath and the system operator $X$ such as the system's canonical
coordinate does not depend on time explicitly. From Eq. (\ref{a52}) the
dissipation matrix element due to the thermal bath is now given by%
\begin{eqnarray}
\mathcal{D}_{mn,m^{\prime }n^{\prime }}^{I} &=&\frac{1}{\hbar ^{2}}\left\{
-\delta _{nn^{\prime }}\sum_{k}X_{mk}X_{km^{\prime }}F^{\ast }(\omega
_{km^{\prime }},t)\right.  \notag \\
&&+X_{mm^{\prime }}X_{n^{\prime }n}\left[ F(\omega _{nn^{\prime
}},t)+F^{\ast }(\omega _{mm^{\prime }},t)\right]  \notag \\
&&-\left. \delta _{mm^{\prime }}\sum_{k}X_{n^{\prime }k}X_{kn}F(\omega
_{kn^{\prime }},t)\right\} ,  \label{a82}
\end{eqnarray}%
where, the superscripts $\mu $ and $\nu $ are omitted for simplicity, $%
F(\omega ,t)$ is given by Eq. (\ref{a51}), and $F^{\ast }(\omega ,t)$ is the
complex conjugate of $F(\omega ,t)$.

If the spectral density of the thermal bath at temperature $T$ is denoted by 
$J_{\varsigma }\left( \omega \right) $, the autocorrelation function $%
\mathcal{J}\left( t\right) $ can be calculated from $J_{\varsigma }\left(
\omega \right) $ by Eq. (\ref{m26-3}). Substituting Eq. (\ref{m26-3}) into
Eq. (\ref{a51}) we obtain%
\begin{equation}
F(\omega ,t)=F_{R}(\omega ,t)+iF_{I}(\omega ,t),  \label{3-7}
\end{equation}%
where, the real part and imaginary part of $F(\omega ,t)$, $F_{R}(\omega ,t)$
and $F_{I}(\omega ,t)$, are given by%
\begin{equation}
F_{\Theta }(\omega ,t)=\frac{1}{2\pi }\int_{-\infty }^{+\infty }J_{\varsigma
}\left( \omega ^{\prime }\right) \lambda _{\Theta }\left( \omega ^{\prime
}+\omega ,t\right) d\omega ^{\prime },  \label{3-8-5}
\end{equation}%
where, $\Theta =R$ and $I$, and $\lambda _{R}\left( \omega ,t\right) =\sin
\left( \omega t\right) /\omega $ and $\lambda _{I}\left( \omega ,t\right) =%
\left[ 1-\cos \left( \omega t\right) \right] /\omega $ are the real part and
imaginary part of $\lambda \left( \omega ,t\right) $ given by 
\begin{equation}
\lambda \left( \omega ,t\right) =\int_{0}^{t}e^{i\omega \tau }d\tau .
\label{3-8-1}
\end{equation}

Substituting Eq. (\ref{3-7}) into Eq. (\ref{a82}) we obtain%
\begin{equation}
\mathcal{D}_{mn,m^{\prime }n^{\prime }}^{I}=\mathcal{R}_{mn,m^{\prime
}n^{\prime }}+i\mathcal{B}_{mn,m^{\prime }n^{\prime }},  \label{3-8-14}
\end{equation}%
where, $\mathcal{R}_{mn,m^{\prime }n^{\prime }}$, which we call damping rate
matrix element, is the rate matrix element related to the change of density
matrix from the value of $\rho _{m^{\prime }n^{\prime }}$ to the value of $%
\rho _{mn}$ \cite{Toutounji05}%
\begin{eqnarray}
\mathcal{R}_{mn,m^{\prime }n^{\prime }}\left( t\right) &=&\frac{1}{\hbar ^{2}%
}\left\{ -\delta _{nn^{\prime }}\sum_{k}X_{mk}X_{km^{\prime }}F_{R}(\omega
_{km^{\prime }},t)\right.  \notag \\
&&+X_{mm^{\prime }}X_{n^{\prime }n}\left[ F_{R}(\omega _{nn^{\prime
}},t)+F_{R}(\omega _{mm^{\prime }},t)\right]  \notag \\
&&-\left. \delta _{mm^{\prime }}\sum_{k}X_{n^{\prime }k}X_{kn}F_{R}(\omega
_{kn^{\prime }},t)\right\} ,  \label{3-8-15}
\end{eqnarray}%
and $\mathcal{B}_{mn,m^{\prime }n^{\prime }}$ is the Lamb shift matrix
element which leads to Lamb shifts of unperturbed energy levels \cite%
{Louisell1973,Breuer2002}%
\begin{eqnarray}
\mathcal{B}_{mn,m^{\prime }n^{\prime }}\left( t\right) &=&\frac{1}{\hbar ^{2}%
}\left\{ \delta _{nn^{\prime }}\sum_{k}X_{mk}X_{km^{\prime }}F_{I}(\omega
_{km^{\prime }},t)\right.  \notag \\
&&+X_{mm^{\prime }}X_{n^{\prime }n}\left[ F_{I}(\omega _{nn^{\prime
}},t)-F_{I}(\omega _{mm^{\prime }},t)\right]  \notag \\
&&-\left. \delta _{mm^{\prime }}\sum_{k}X_{n^{\prime }k}X_{kn}F_{I}(\omega
_{kn^{\prime }},t)\right\} .  \label{3-8-16}
\end{eqnarray}%
They satisfy the symmetric relations%
\begin{equation}
\mathcal{R}_{nm,n^{\prime }m^{\prime }}\left( t\right) =\mathcal{R}%
_{mn,m^{\prime }n^{\prime }}\left( t\right) ,  \label{3-35}
\end{equation}%
and%
\begin{equation}
\mathcal{B}_{nm,n^{\prime }m^{\prime }}\left( t\right) =-\mathcal{B}%
_{mn,m^{\prime }n^{\prime }}\left( t\right) .  \label{3-36}
\end{equation}%
From Eq. (\ref{3-36}) $\mathcal{B}_{mm,nn}\left( t\right) =0$.

\subsection{Steady Lamb shift matrix and damping rate matrix at a long time
limit}

In general, the damping rate matrix and Lamb shift matrix are time-dependent
because both $F_{R}(\omega ,t)$ and $F_{I}(\omega ,t)$ depend on time.
However, in the investigation of decoherence of a qubit, one is only
interested in the behavior of the qubit after a sufficiently long time. In
this case, the damping rate matrix and Lamb shift matrix can be well
approximated by steady ones.

In fact, the autocorrelation function of a thermal bath impacts the system
only in some time interval $t_{c}$ which is called the correlation time \cite%
{Louisell1973}. As long as the upper limit in the integral of Eq. (\ref%
{3-8-1}) $t\gg t_{c}$ it may be extended to the infinity with very little
error.

Applying the integral 
\begin{equation}
\int_{0}^{\infty }e^{\pm i\omega \tau }d\tau =\pi \delta \left( \omega
\right) \pm i\mathcal{P}\frac{1}{\omega },  \label{3-37}
\end{equation}
where $\mathcal{P}$ represents the Cauchy principal value of the integral,
to Eq. (\ref{3-8-1}) one has%
\begin{eqnarray}
\lambda _{R}\left( \omega ,\infty \right) &=&\pi \delta \left( \omega
\right) ,  \label{3-8-12} \\
\lambda _{I}\left( \omega ,\infty \right) &=&\mathcal{P}\frac{1}{\omega }.
\label{3-8-13}
\end{eqnarray}%
Substituting Eqs. (\ref{3-8-12}) and (\ref{3-8-13}) into Eq. (\ref{3-8-5}),
one obtains%
\begin{eqnarray}
f_{R}(\omega ) &=&F_{R}(\omega ,\infty )=\frac{1}{2}J_{\varsigma }\left(
-\omega \right) ,  \label{3-8-80} \\
f_{I}(\omega ) &=&F_{I}(\omega ,\infty )=\frac{1}{2\pi }\mathcal{P}%
\int_{-\infty }^{+\infty }\frac{J_{\varsigma }\left( \omega ^{\prime
}\right) }{\omega ^{\prime }+\omega }d\omega ^{\prime }.  \label{3-8-81}
\end{eqnarray}

In Eqs. (\ref{3-8-15}) and (\ref{3-8-16}), replacing $F_{R}$ with $f_{R}$
and $F_{I}$ with $f_{I}$, we obtain steady damping rate matrix element $%
R_{mn,m^{\prime }n^{\prime }}$ and Lamb shift matrix element $%
B_{mn,m^{\prime }n^{\prime }}$. They are given by 
\begin{eqnarray}
R_{mn,m^{\prime }n^{\prime }} &=&\frac{1}{\hbar ^{2}}\left[ -\delta
_{nn^{\prime }}\sum_{k}X_{mk}X_{km^{\prime }}f_{R}(\omega _{km^{\prime
}})\right.  \notag \\
&&+X_{mm^{\prime }}X_{n^{\prime }n}\left[ f_{R}(\omega _{nn^{\prime
}})+f_{R}(\omega _{mm^{\prime }})\right]  \notag \\
&&-\left. \delta _{mm^{\prime }}\sum_{k}X_{n^{\prime }k}X_{kn}f_{R}(\omega
_{kn^{\prime }})\right] ,  \label{3-11-1}
\end{eqnarray}%
and%
\begin{eqnarray}
B_{mn,m^{\prime }n^{\prime }} &=&\frac{1}{\hbar ^{2}}\left[ \delta
_{nn^{\prime }}\sum_{k}X_{mk}X_{km^{\prime }}f_{I}(\omega _{km^{\prime
}})\right.  \notag \\
&&+X_{mm^{\prime }}X_{n^{\prime }n}\left[ f_{I}\left( \omega _{nn^{\prime
}}\right) -f_{I}\left( \omega _{mm^{\prime }}\right) \right]  \notag \\
&&-\left. \delta _{mm^{\prime }}\sum_{k}X_{n^{\prime }k}X_{kn}f_{I}(\omega
_{kn^{\prime }})\right] ,  \label{3-11-2}
\end{eqnarray}%
respectively. The $R_{mn,m^{\prime }n^{\prime }}$ and $B_{mn,m^{\prime
}n^{\prime }}$ have the same symmetric relations as $\mathcal{R}%
_{mn,m^{\prime }n^{\prime }}$ and $\mathcal{B}_{mn,m^{\prime }n^{\prime }}$
given by Eqs. (\ref{3-35}) and (\ref{3-36}).

It is shown from Eq. (\ref{3-8-14}) that the thermal bath affects the system
via the damping rate matrix and Lamb shift matrix. The damping rate matrix
elements represent decay rates in special cases. For instance, $R_{mm,mm}$
characterizes the rate of population change in the state $m$, $R_{nn,mm}$
describes the rate of population transfer from the states $m$ to $n$, and $%
R_{mn,mn}$ represents the dephasing (coherence decay between the states $m$
to $n$) rate of the off-diagonal elements of $\rho _{mn}$. In contrast, the
Lamb shift matrix element such as $B_{mn,mm}$ represents the well-known Lamb
shift of the state $m$ induced by the thermal bath. Hence, the damping rate
matrix makes the system relax and decohere while the Lamb shift matrix makes
the energy levels shift. In addition, the symmetries of the Lamb shift
matrix given by Eq. (\ref{3-36}) are the same as those of the matrix for
driving fields given by Eq. (\ref{m42-1}). In particular, when substituting
Eq. (\ref{3-8-14}) into Eq.\ (\ref{n3}) the Lamb shift matrix can be
incorporated into the matrix $\mathcal{L}_{mn,m^{\prime }n^{\prime }}^{F}$
and leads to the renormalization of the quantum system Hamiltonian. \cite%
{Breuer2002} Thus the effect of the Lamb shift matrix is analogous to an
extra\ field. In the case of weak damping, the Lamb shift matrix is very
small compared to the driving field and thus is neglected hereafter.

\subsection{Spectral density, spontaneous decay, stimulated transition, and
detailed balance}

The fluctuation of the thermal bath at temperature $T$ is characterized by a
spectral density. The spectral density is computed from the real part of the
frequency-dependent damping coefficient $\gamma _{R}\left( \omega \right) $
in terms of the quantum fluctuation-dissipation theorem (see the Appendix B
for details). \cite{Weiss1999,Devoret97} For the interaction given by Eq. (%
\ref{m20}), the fluctuation force $\xi \left( t\right) $ is proportional to
the fluctuation force $\varsigma \left( t\right) $ by%
\begin{equation}
\xi \left( t\right) =\Lambda \varsigma \left( t\right) ,  \label{2-11}
\end{equation}%
where,%
\begin{equation}
\Lambda =-\frac{\partial X}{\partial q},  \label{2-11-1}
\end{equation}%
and $q$ is the coordinate operator of the system. For the weak damping
considered here the interaction given by Eq. (\ref{m20}) is a linear
function of system coordinate operator and thus $\Lambda $ is a constant.

For the thermal bath at temperature $T$, the spectral density $J_{\xi
}\left( \omega \right) $ of the fluctuation force $\xi \left( t\right) $ is
given by Eq. (\ref{A3}). From Eqs. (\ref{A3}) and (\ref{2-11}), the spectral
density $J_{\varsigma }\left( \omega \right) $ of the fluctuation force $%
\varsigma \left( t\right) $ is given by 
\begin{equation}
J_{\varsigma }\left( \omega \right) =\frac{J_{\xi }\left( \omega \right) }{%
\Lambda ^{2}}=\frac{M\hbar \omega }{\Lambda ^{2}}\gamma _{R}\left( \omega
\right) \left[ 1+\coth \left( \frac{\hbar \omega }{2k_{B}T}\right) \right] .
\label{2-6-0}
\end{equation}%
Substituting Eq. (\ref{2-6-0}) into Eq. (\ref{m26-3}), we obtain the
autocorrelation function of the fluctuation force $\varsigma \left( t\right) 
$ 
\begin{eqnarray}
\mathcal{J}(t) &=&\frac{M\hbar }{2\pi \Lambda ^{2}}\int_{-\infty }^{+\infty
}\omega \gamma _{R}\left( \omega \right) \left[ 1+\coth \left( \frac{\hbar
\omega }{2k_{B}T}\right) \right]  \notag \\
&&\times \exp \left( i\omega t\right) d\omega .  \label{2-6}
\end{eqnarray}%
In general, the autocorrelation function given by Eq. (\ref{2-6}) is complex
and irreversible since the integrand does not have a definite symmetry with
respect to $\omega $. This result is totally different from that for a
classical system, for which the autocorrelation function is real and
reversible. As has been demonstrated, the irreversibility of the
autocorrelation function results from the non-Hermitian property and
noncommutable property of the thermal bath operators.

Substituting Eqs. (\ref{2-6-0}) and (\ref{3-8-80}) into Eq. (\ref{3-11-1}),
we obtain for $m\neq n$ the rate of population transfer from the state $n$
to $m$%
\begin{eqnarray}
R_{mm,nn} &=&-\frac{M\left\vert X_{mn}\right\vert ^{2}}{\hbar \Lambda ^{2}}%
\omega _{mn}\gamma _{R}(-\omega _{mn})  \notag \\
&&\times \left[ 1-\coth \left( \frac{\hbar \omega _{mn}}{2k_{B}T}\right) %
\right] .  \label{3-13}
\end{eqnarray}%
From Eqs. (\ref{3-11-1}) and (\ref{3-13}) we also obtain the rate of
population change in the state $n$%
\begin{equation}
R_{nn,nn}=-\sum_{m\left( \neq n\right) }R_{mm,nn}.  \label{3-17}
\end{equation}%
This indicates that $R_{nn,nn}$ is a sum of the rates of population transfer
from the state $n$ to all the other states $m$. This result is identical
with that of atomic systems. \cite{Louisell1973}

\subsubsection{Spontaneous decay}

Let us consider a pair of states denoted by $m$ and $n$. Suppose the state $%
m $ has lower energy than the state $n$. The transition frequency is $\omega
_{nm}=\left( E_{n}-E_{m}\right) /\hbar >0$. If the system is in the higher
energy state initially, it can decay without presence of any external field
due to the stimulation of thermal fluctuation of the thermal bath. This
process is the so-called spontaneous decay. During this process the system
transits from the higher energy state $n$ to the lower energy state $m$ with
emission of radiation. From Eq. (\ref{3-13}) the spontaneous decay rate $%
\Gamma _{mn}^{SP}$ is given by%
\begin{eqnarray}
\Gamma _{mn}^{SP} &=&\frac{M\left\vert X_{mn}\right\vert ^{2}}{\hbar \Lambda
^{2}}\omega _{nm}\gamma _{R}(\omega _{nm})  \notag \\
&&\times \left[ 1+\coth \left( \frac{\hbar \omega _{nm}}{2k_{B}T}\right) %
\right] .  \label{3-14}
\end{eqnarray}%
It is proportional to the transition matrix element $X_{mn}$, energy level
spacing $\omega _{nm}$, and damping coefficient $\gamma _{R}(\omega _{nm})$.
It also depends on the temperature via the factor $\coth \left( \hbar \omega
_{nm}/2k_{B}T\right) $.

\subsubsection{Stimulated transition}

If the system is in the lower energy state initially it may occur an inverse
process with respect to the spontaneous decay under the stimulation of
thermal fluctuation, making the system transit from the lower energy state $%
m $ to the higher energy state $n$. This process is the so-called stimulated
transition. From Eq. (\ref{3-13}) the stimulated transition rate $\Gamma
_{nm}^{ST}$ is given by%
\begin{eqnarray}
\Gamma _{nm}^{ST} &=&-\frac{M\left\vert X_{mn}\right\vert ^{2}}{\hbar
\Lambda ^{2}}\omega _{nm}\gamma _{R}(-\omega _{nm})  \notag \\
&&\times \left[ 1-\coth \left( \frac{\hbar \omega _{nm}}{2k_{B}T}\right) %
\right] .  \label{3-15}
\end{eqnarray}

\subsubsection{Detailed balance}

In general, $\gamma _{R}\left( \omega \right) $ is an even function of $%
\omega $, i.e., $\gamma _{R}(-\omega )=\gamma _{R}(\omega )$. Using Eqs. (%
\ref{3-14}) and (\ref{3-15}) one has%
\begin{equation}
\frac{\Gamma _{mn}^{SP}}{\Gamma _{nm}^{ST}}=\exp \left( \frac{E_{n}-E_{m}}{%
k_{B}T}\right) .  \label{3-16}
\end{equation}%
Thus the ratio of the spontaneous decay rate to the stimulated transition
rate satisfies the detailed balance principle.

\section{Decoherence of a dissipative two-level system}

An ideal qubit is a two-level system and an ideal qubit interacting with a
thermal bath is equivalent to a dissipative two-level system (DTLS). For a
DTLS, the master equation (\ref{n3}) can be replaced by the Bloch equation. 
\cite{zhou2006field_effect} In the rotating-wave approximation (RWA),
analytical solutions of the Bloch equation can be obtained, from which
analytical expressions of characteristic (relaxation, decoherence, and
dephasing) times of the DTLS can be derived. In this section, we present
some important analytical results without derivation. The details about
these analytical results can be found in our previous paper. \cite%
{zhou2006field_effect}

\subsection{Relaxation and decoherence times of the DTLS in the absence of
driving fields}

In the absence of driving fields, the relaxation and decoherence times of
the DTLS are given by%
\begin{equation}
T_{1}=\kappa _{1}^{-1},  \label{x-80}
\end{equation}%
and 
\begin{equation}
T_{2}=\kappa _{2}^{-1},  \label{x-81}
\end{equation}%
respectively, where, $\kappa _{1}$ is the relaxation rate of the DTLS in
free decay given by 
\begin{equation}
\kappa _{1}=R_{22,11}+R_{11,22},  \label{i-4-2}
\end{equation}%
and $\kappa _{2}$ is the decoherence rate of the the DTLS in free decay
given by%
\begin{equation}
\kappa _{2}=-R_{12,12}.  \label{m-8-2-1}
\end{equation}%
These results accord with those obtained by others. \cite{Burkard2004} In
general, due to dephasing $T_{2}<2T_{1}$. \cite{Falci05} The dephasing time $%
T_{\varphi }$ can be calculated from $T_{1}$ and $T_{2}$ by \cite%
{Burkard2004}%
\begin{equation}
\frac{1}{T_{\varphi }}=\frac{1}{T_{2}}-\frac{1}{2T_{1}}.  \label{x-81-1}
\end{equation}

\subsection{Relaxation and decoherence times of the DTLS in the presence of
a resonant ac driving field}

For an DTLS resonantly driven by an ac field, multiple relaxation and
decoherence times are required to completely describe time evolution of
population and coherence of the qubit. \cite{zhou2006field_effect}
Particularly, intrinsic and field-induced decoherence times are necessitated
to characterize the decoherence. In the underdamped regime, the most
important regime for the driven qubit, the relaxation time $\widetilde{T}%
_{1} $ is given by%
\begin{equation}
\widetilde{T}_{1}=\Gamma ^{-1},  \label{x-82}
\end{equation}%
where, the tilt "$\sim $" is used to denote the characteristic times of the
resonantly driven qubit and $\Gamma $ is the relaxation rate of the driven
DTLS given by%
\begin{equation}
\Gamma =\frac{\kappa _{1}+\kappa _{2}}{2}.  \label{x-82-1}
\end{equation}%
The intrinsic decoherence time $\widetilde{T}_{21}$ and field-induced
decoherence time $\widetilde{T}_{22}$ are given by%
\begin{equation}
\widetilde{T}_{21}=\kappa _{2}^{-1}=T_{2},  \label{x-84}
\end{equation}%
and%
\begin{equation}
\widetilde{T}_{22}=\Gamma ^{-1}=\widetilde{T}_{1},  \label{x-84-1}
\end{equation}%
respectively. It is shown that due to the effect of driving fields the
characteristic times of the driven DTLS are different from those of the DTLS
in free decay. The characteristic times of the driven DTLS are independent
of the field strength in the case of weak driving fields. If the initial
state of the driven DTLS is an eigenstate such as the ground state, the
intrinsic decoherence vanishes and the decoherence is completely
characterized by the field-induced decoherence. In this case, the
decoherence rate of the driven DTLS is equal to its relaxation rate as shown
by Eq. (\ref{x-84-1}). Therefore, the escape rate of quantum phase
information from the system equals to the rate of energy flowing from the
system to the environment. This result is identical with that obtained from
the non-Markovian approach with inclusion of both bath and qubit dynamics. 
\cite{Shresta05}

From Eqs. (\ref{x-80}), (\ref{x-81}), (\ref{x-84}), and (\ref{x-84-1}) we
obtain%
\begin{equation}
\frac{1}{\widetilde{T}_{22}}=\frac{1}{\widetilde{T}_{1}}=\frac{1}{2T_{1}}+%
\frac{1}{2T_{2}}=\frac{3}{4T_{1}}+\frac{1}{2T_{\varphi }}.  \label{x-86}
\end{equation}%
It shown that $\widetilde{T}_{22}=\widetilde{T}_{1}$, $\min \left(
T_{1},T_{2}\right) \leq \widetilde{T}_{1}\left( \widetilde{T}_{22}\right)
\leq \max \left( T_{1},T_{2}\right) $, and $\widetilde{T}_{1}\left( 
\widetilde{T}_{22}\right) <\min \left( 4T_{1}/3,2T_{\varphi }\right) $.
These results are similar to those obtained by others \cite%
{Anastopoulos2000,Kosugi05} and also agrees, within the experimental
uncertainties, with recent experimental results. \cite{Ithier05}

\section{Decoherence of a 2D SQUID flux qubit coupled to control and readout
circuits}

A number of effects can destroy coherence of a SQUID qubit, \cite{Tian2000}
of which the fluctuation of external circuits used for control and readout
of the qubit is one of the important sources of decoherence. \cite%
{Makhlin2001} In this section, we investigate decoherence of a 2D SQUID flux
qubit due to coupling with control and readout circuits and effect of
driving fields on the decoherence.

\subsection{Hamiltonian of the microwave-driven 2D SQUID flux qubit}

As shown in FIG. 1\textrm{, }a 2D SQUID flux qubit is a variable barrier rf
SQUID in which the single Josephson junction in an ordinary rf SQUID is
replaced by a low inductance dc SQUID. \cite{Lishaoxiong06} Suppose that the
inductance of the superconducting loop of the rf SQUID is $L$, the critical
current is $I_{c}$, the total magnetic flux enclosed in the rf SQUID loop is 
$\Phi $, the shunt capacitance of each Josephson junction in the dc SQUID is 
$C$, the critical currents are $I_{c1}$ and $I_{c2}$, and the total magnetic
flux enclosed in the dc SQUID loop is $\Phi _{dc}$. The Hamiltonian of the
2D SQUID qubit can be written as 
\begin{equation}
H_{S}\left( x,y\right) =\frac{p_{x}^{2}}{2m_{x}}+\frac{p_{y}^{2}}{2m_{y}}%
+V\left( x,y\right) ,  \label{b0}
\end{equation}%
where, $m_{x}=2C\Phi _{0}^{2}$ and $m_{y}=C\Phi _{0}^{2}/2$ are the masses
of the first and second modes, $\Phi _{0}\equiv h/2e$ and $e$ are the flux
quantum and elementary charge, $x=\Phi /\Phi _{0}$ and $y=\Phi _{dc}/\Phi
_{0}$ are the canonical coordinates of the 2D SQUID qubit, $p_{x}=-i\hbar
\partial /\partial x$ and $p_{y}=-i\hbar \partial /\partial y$ are the
canonical momenta conjugate to $x$ and $y$, and $V(x,y)$ is the potential
energy given by \cite{Han1992}%
\begin{eqnarray}
V\left( x,y\right)  &=&\frac{\Phi _{0}^{2}}{L}\left[ \frac{1}{2}\left(
x-x_{e}\right) ^{2}+\frac{g}{2}(y-y_{e})^{2}\right.   \notag \\
&&-\frac{\beta _{L}}{4\pi ^{2}}\cos \left( 2\pi x\right) \cos \left( \pi
y\right)   \notag \\
&&\left. +\frac{\delta \beta _{L}}{4\pi ^{2}}\sin \left( 2\pi x\right) \sin
\left( \pi y\right) \right] .  \label{b1}
\end{eqnarray}%
Here, $g\equiv L/2l$ is the ratio of the inductances of the rf SQUID and dc
SQUID, $\beta _{L}\equiv 2\pi LI_{c}/\Phi _{0}$, $\delta \beta _{L}\equiv
2\pi L\left( I_{c2}-I_{c1}\right) /\Phi _{0}$, $x_{e}$ and $y_{e}$ are the
fluxes applied to the rf SQUID and dc SQUID in the unit of $\Phi _{0}$. The
Josephson coupling energy of the rf SQUID is $E_{J}=\hbar I_{c}/2e=m\omega
_{LC}^{2}\beta _{L}/4\pi ^{2}$, where $\omega _{LC}=1/\sqrt{LC}$ is the
characteristic frequency of the 2D SQUID qubit. The contour of the potential
energy of the 2D SQUID qubit used in recent experiment \cite{Lishaoxiong06}
is plotted in FIG. 2, where the parameters of the 2D SQUID qubit are $L=205$
pH, $C=32.5$ fF, $g=17.0$, $\beta _{L}=3.7$, $\delta \beta _{L}=0$, $%
x_{e}=0.4991$, and $y_{e}=0.387$.

The spectroscopic properties of the 2D SQUID qubit can be obtained by
solving the eigenvalue equation of Hamiltonian $H_{S}$. \cite%
{Zhou2002,Zhou2004} In FIG. 3 and FIG. 4, we plot the energy levels and
transition matrix elements versus the flux applied to the rf SQUID $x_{e}$,
respectively. When $x_{e}=0.4991$, at the position of the arrow, $\Delta
E_{31}=E_{3}-E_{1}=0.259\omega _{LC}=15.95$ GHz and $\left\vert
x_{21}/x_{32}\right\vert =0.262$, where $\omega _{LC}=3.874\times 10^{11}$
rad/s. These results are in very good agreement with the experimental
results. \cite{Lishaoxiong06} Note that $x_{e}$ and $y_{e}$ determine the
energy bias and tunnel splitting, respectively. Thus the spectroscopic
properties of the 2D SQUID qubit can be varied \textit{in situ} by adjusting 
$x_{e}$ and $y_{e}$.

In order to manipulate the qubit's states for gate operations, a microwave
pulse is applied to the 2D SQUID qubit through the first mode. If the
interaction of the microwave field and external circuits is neglected, the
Hamiltonian of the microwave-driven 2D SQUID qubit coupled to the external
circuits is given by%
\begin{equation}
H\left( x,t\right) =H_{S}\left( x,y\right) +H_{F}\left( x,t\right)
+H_{I}\left( x,y,t\right) ,  \label{b1-1}
\end{equation}%
where, $H_{F}\left( x,t\right) $ is the interaction of the SQUID qubit and
the microwave field and $H_{I}\left( x,y,t\right) $ is the interaction of
the SQUID qubit and the external circuits (thermal bath).

If $\phi \left( t\right) $ is the normalized flux from the microwave field
coupled to the SQUID qubit, which is taken to be%
\begin{equation}
\phi (t)=\phi _{\mu }\cos \left( \omega _{\mu }t\right) ,  \label{b4-1}
\end{equation}%
where, $\phi _{\mu }$ and $\omega _{\mu }$ are the field strength and
frequency, respectively, then $H_{F}\left( x,t\right) $ is given by \cite%
{Zhou2002}%
\begin{equation}
H_{F}\left( x,t\right) =\frac{\Phi _{0}^{2}}{2L}\phi \left[ \phi -2\left(
x-x_{e}\right) \right] .  \label{b4}
\end{equation}

In general, the external circuits are coupled to the 2D SQUID qubit through
both modes. However, the coupling through the second mode is negligibly
small compared to that through the first mode and will be neglected. In the
case of weak coupling, the interaction of the 2D SQUID qubit and the
external circuits is a linear function of $x$ and can be expressed by \cite%
{Makhlin2001,Xu2005}%
\begin{equation}
H_{I}(x,t)=x\varsigma \left( t\right) ,  \label{z-5}
\end{equation}%
where, $\varsigma \left( t\right) $ is the fluctuation force of the external
circuits. If the coupling between the control and readout circuits is
neglected, $\varsigma \left( t\right) $ can be decomposed into two parts as%
\begin{equation}
\varsigma \left( t\right) =\varsigma _{x}\left( t\right) +\varsigma _{m}(t),
\label{z-5-1}
\end{equation}%
where, $\varsigma _{x}\left( t\right) $ and $\varsigma _{m}(t)$ are the
fluctuation forces of the control and readout circuits, respectively.

\subsection{Spectral densities of the control and readout circuits}

As has been demonstrated, in the case of weak damping, the effect of a
thermal bath on a quantum system can be characterized by a spectral density. 
\cite{Leggett87,Devoret97} The spectral density can be computed from the
frequency-dependent damping coefficient of the quantum Langevin equation
(see Appendix B). For the 2D SQUID flux qubit, the quantum Langevin equation
is given by \cite{Devoret97}%
\begin{equation}
C\frac{d^{2}\Phi (t)}{dt^{2}}+\int_{t_{0}}^{t}dt^{\prime }Y\left(
t-t^{\prime }\right) \frac{d\Phi (t^{\prime })}{dt^{\prime }}+\frac{dV\left(
\Phi \right) }{d\Phi }=\xi _{\Phi }\left( t\right) ,  \label{y6}
\end{equation}%
where, the damping coefficient $Y(t)$ is the equivalent admittance of the
external circuits, $V\left( \Phi \right) $ is the potential energy applied
to the SQUID qubit, and $\xi _{\Phi }\left( t\right) $ is the fluctuation
force of the external circuits.

By comparison of Eq. (\ref{y6}) with Eq. (\ref{A1}), one has $q\rightarrow
\Phi $, $M\rightarrow C$, $\gamma \rightarrow Y/C$, and $\xi \rightarrow \xi
_{\Phi }$. Applying these relations to Eq. (\ref{A3}), we obtain the
spectral density $J\left( \omega \right) $ of the fluctuation force $\xi
_{\Phi }\left( t\right) $%
\begin{equation}
J\left( \omega \right) =\hbar \omega Y_{R}(\omega )\left[ 1+\coth \left( 
\frac{\hbar \omega }{2k_{B}T}\right) \right] ,  \label{3-2}
\end{equation}%
where, $T$ is the temperature of the external circuits and $Y_{R}(\omega )$
is the real part of the frequency-dependent equivalent admittance $Y(\omega )
$ of the external circuits. $Y(\omega )$ can be calculated from $Y(t)$ by
Fourier transform. It can also be calculated directly from the external
circuits using circuit equations in the frequency domain. For the control
and readout circuits given in FIG. 1, $Y(\omega )$ has been derived (see
Appendix C for details) and $Y_{R}(\omega )$ is given by%
\begin{equation}
Y_{R}(\omega )=Y_{xR}(\omega )+Y_{mR}(\omega ),  \label{i3-2-1}
\end{equation}%
where, $Y_{xR}(\omega )$ and $Y_{mR}(\omega )$ are the real parts of the
frequency-dependent admittances of the control circuit and readout circuit,
respectively. They are given by%
\begin{equation}
Y_{xR}\left( \omega \right) =\frac{F_{x}\left( \omega \right) }{\omega
^{2}+G_{x}\left( \omega \right) },  \label{i3-2-2}
\end{equation}%
and%
\begin{equation}
Y_{mR}\left( \omega \right) =\frac{1}{F_{m}\left( \omega \right) \left[
1+G_{m}\left( \omega \right) \omega ^{2}\right] },  \label{i3-2-3}
\end{equation}%
respectively, where, $F_{x}\left( \omega \right) $, $G_{x}\left( \omega
\right) $, $F_{m}\left( \omega \right) $, and $G_{m}\left( \omega \right) $
are given by Eqs. (\ref{A-a10}), (\ref{A-a11}), (\ref{x-9}), and (\ref{x-10}%
) in Appendix C. Substituting Eq. (\ref{i3-2-1}) into Eq. (\ref{3-2}), one
has%
\begin{equation}
J\left( \omega \right) =J_{x}\left( \omega \right) +J_{m}\left( \omega
\right) ,  \label{i3-2-4}
\end{equation}%
where, $J_{x}\left( \omega \right) $ and $J_{m}\left( \omega \right) $ are
the spectral densities of the control circuit and readout circuit,
respectively. They can be calculated using Eq. (\ref{3-2}) with $%
Y_{R}(\omega )$ replaced by $Y_{xR}(\omega )$ and $Y_{mR}(\omega )$,
respectively.

In FIG. 5 we plot the spectral densities $J_{x}\left( \omega \right) $, $%
J_{m}\left( \omega \right) $, and $J\left( \omega \right) $ versus the
frequency $\omega $ for the control and readout circuits of the 2D SQUID
qubit used in the experiment. \cite{Lishaoxiong06} The parameters of the
control circuit are $L_{x}=100$ pH, $C_{x}=25$ pF, $R_{x}=70$ $\Omega $, and 
$R_{x0}=1.0\times 10^{3}$ $\Omega $. The parameters of the readout circuit
are $L_{10}=L_{20}=20$ pH, $L_{J1}=100$ pH, $L_{J2}=550$ pH, $C_{m}=20$ pF, $%
R_{m}=70$ $\Omega $, and $R_{m0}=2.0\times 10^{4}$ $\Omega $. The mutual
inductances $M_{x}=1.0$ pH and $M_{m}=3.3$ pH, and the temperature $T=30$
mK. It is shown that the spectral density $J\left( \omega \right) $ (also $%
J_{x}\left( \omega \right) $ and $J_{m}\left( \omega \right) $) reaches the
maximum at $\omega \simeq 1.69\omega _{LC}$. At low frequency, the spectral
density $J\left( \omega \right) $ (also $J_{x}\left( \omega \right) $ and $%
J_{m}\left( \omega \right) $) approaches a constant value, in particular $%
J\left( 0\right) \neq 0$. As will be demonstrated, the dephasing time $%
T_{\varphi }$ is finite if $J\left( 0\right) \neq 0$. Thus the control and
readout circuits of the 2D SQUID qubit induce both relaxation and dephasing.

\subsection{Damping rate matrix}

From Eq. (\ref{z-5}), the fluctuation force $\xi _{\Phi }\left( t\right) $
is given by%
\begin{equation}
\xi _{\Phi }\left( t\right) =-\frac{\partial H_{I}(x,t)}{\partial \Phi }=-%
\frac{1}{\Phi _{0}}\varsigma \left( t\right) .  \label{ii-1}
\end{equation}%
This equation provides a relation between the fluctuation force $\varsigma
\left( t\right) $ and $\xi _{\Phi }\left( t\right) $ of the external
circuits. Compared Eq. (\ref{ii-1}) with Eq. (\ref{2-11}), one has $\Lambda
=-1/\Phi _{0}$. Using Eq. (\ref{2-6-0}), the spectral density $J_{\varsigma
}\left( \omega \right) $ of the fluctuation force $\varsigma \left( t\right) 
$ can be calculated from the spectral density $J\left( \omega \right) $ of
the fluctuation force $\xi _{\Phi }\left( t\right) $ by%
\begin{equation}
J_{\varsigma }\left( \omega \right) =\Phi _{0}^{2}J\left( \omega \right) .
\label{3-3}
\end{equation}%
For the 2D SQUID qubit, $X_{1}=x$ and $X_{2}=y$. Substituting Eqs. (\ref%
{3-8-80}) and (\ref{3-3}) into Eq. (\ref{3-11-1}) we obtain the steady
damping rate matrix element $R_{mn,m^{\prime }n^{\prime }}$ of the 2D SQUID
qubit

\begin{eqnarray}
R_{mn,m^{\prime }n^{\prime }} &=&\frac{\Phi _{0}^{2}}{2\hbar ^{2}}\left[
-\delta _{nn^{\prime }}\sum_{k}x_{mk}x_{km^{\prime }}J(\omega _{m^{\prime
}k})\right.  \notag \\
&&+x_{mm^{\prime }}x_{n^{\prime }n}\left[ J(\omega _{n^{\prime }n})+J(\omega
_{m^{\prime }m})\right]  \notag \\
&&-\left. \delta _{mm^{\prime }}\sum_{k}x_{n^{\prime }k}x_{kn}J(\omega
_{n^{\prime }k})\right] .  \label{3-3-1}
\end{eqnarray}%
It is a function of transition matrix elements and spectral densities.

If only considering the two computational levels, the 2D SQUID qubit is
equivalent to a DTLS and its characteristic (relaxation, decoherence, and
dephasing) times can be calculated analytically. Substituting Eq. (\ref%
{3-3-1}) into Eqs. (\ref{i-4-2}) and (\ref{m-8-2-1}), then Eqs. (\ref{i-4-2}%
) and (\ref{m-8-2-1}) into Eqs. (\ref{x-80}), (\ref{x-81}), and (\ref{x-81-1}%
), one has%
\begin{eqnarray}
T_{1}^{-1} &=&\frac{\pi ^{2}}{e^{2}}\left\vert x_{12}\right\vert ^{2}\left[
J\left( \omega _{21}\right) +J\left( \omega _{12}\right) \right] ,
\label{x-81-1-0} \\
T_{2}^{-1} &=&\frac{1}{2T_{1}}+\frac{\pi ^{2}}{2e^{2}}\left(
x_{11}-x_{22}\right) ^{2}J\left( 0\right) ,  \label{x-81-1-1}
\end{eqnarray}%
and%
\begin{equation}
T_{\varphi }^{-1}=\frac{\pi ^{2}}{2e^{2}}\left( x_{11}-x_{22}\right)
^{2}J\left( 0\right) .  \label{x-81-1-2}
\end{equation}%
It is shown that the relaxation and dephasing rates are determined by the
spectral densities at transition frequency $\omega =\left\vert \omega
_{21}\right\vert $ and low frequency $\omega =0$, respectively. \cite%
{Burkard2004,Makhlin04,Ithier05,Bertet06,Berkley03-1,Xu2005} The relaxation
rate is proportional to the modulus square of transition matrix element $%
\left\vert x_{12}\right\vert ^{2}$, while the dephasing rate is proportional
to the squared difference of average coordinates of the two states $\left(
x_{11}-x_{22}\right) ^{2}$. For a qubit having $\left( x_{11}-x_{22}\right)
=0$, the dephasing is completely suppressed. For the 2D SQUID qubit
considered here, both $J\left( 0\right) $ and $\left( x_{11}-x_{22}\right)
^{2}$ are not zero. Thus the control and readout circuits will induce phase
relaxation.

For the 2D SQUID flux qubit, $Y_{xR}\left( \omega \right) $, $Y_{mR}\left(
\omega \right) $, and thus $Y_{R}\left( \omega \right) $ are even functions
of $\omega $. Substituting Eq. (\ref{3-2}) into Eqs. (\ref{x-81-1-0}) to (%
\ref{x-81-1-2}), the analytical expressions of the characteristic times are
simplified to%
\begin{eqnarray}
T_{1}^{-1} &=&\frac{2\pi ^{2}}{e^{2}}\hbar \omega _{21}\left\vert
x_{12}\right\vert ^{2}Y_{R}\left( \omega _{21}\right) \coth \left( \frac{%
\hbar \omega _{21}}{2k_{B}T}\right) ,  \label{x-81-1-3-1} \\
T_{2}^{-1} &=&\frac{1}{2T_{1}}+\frac{\pi ^{2}}{e^{2}}k_{B}T\left(
x_{11}-x_{22}\right) ^{2}Y_{R}\left( 0\right) ,  \label{x-81-3}
\end{eqnarray}%
and%
\begin{equation}
T_{\varphi }^{-1}=\frac{\pi ^{2}}{e^{2}}k_{B}T\left( x_{11}-x_{22}\right)
^{2}Y_{R}\left( 0\right) .  \label{x-81-1-4}
\end{equation}%
Now the relaxation rate is dominated by the external circuits' admittance at
transition frequency $\omega _{21}$ while the dephasing rate by the
admittance at $\omega =0$. Furthermore, the dephasing rate is proportional
to the temperature of the thermal bath. Hence at low temperature the
dominating source of decoherence is relaxation while at high temperature the
main source of decoherence is dephasing. These results agree with those
obtained by others. \cite{Leggett87,Tian2002,Burkard2004}

Using Eq. (\ref{3-14}), the general form of spontaneous decay rate, we
obtain the spontaneous decay rate of the 2D SQUID flux qubit%
\begin{eqnarray}
\Gamma _{mn}^{SP} &=&\frac{2\pi }{\hbar }R_{Q}\left\vert x_{mn}\right\vert
^{2}\left( E_{n}-E_{m}\right) Y_{R}(\omega _{nm})  \notag \\
&&\times \left[ 1+\coth \left( \frac{E_{n}-E_{m}}{2k_{B}T}\right) \right] ,
\label{3-5-1}
\end{eqnarray}%
where, $R_{Q}$ $=h/4e^{2}$ is the resistance quantum and $\omega
_{nm}=\left( E_{n}-E_{m}\right) /\hbar >0$. If the equivalent impedance is a
resistance $R$, then $Y_{R}(\omega _{nm})=1/R$ and the spontaneous decay
rate is the same as that given by others with different method. \cite%
{Larkin86,Han2001}

\subsection{Numerical method}

In the Hilbert space spanned by the eigenstates of the qubit's Hamiltonian,
the density matrix is governed by the master equation (\ref{n3}). This
equation can be rewritten in matrix form as%
\begin{equation}
\frac{d\mathbf{\rho }}{dt}=-i\mathbf{L\rho }+\mathbf{R\rho },  \label{o-1}
\end{equation}%
where, $\mathbf{\rho }=\left\{ \rho _{\mu }\right\} $ is the density matrix, 
$\mathbf{L}=\left\{ \mathcal{L}_{\mu \mu ^{\prime }}^{S}+\mathcal{L}_{\mu
\mu ^{\prime }}^{F}\right\} $ is the matrix of Liouville superoperators, $%
\mathbf{R}=\left\{ R_{\mu \mu ^{\prime }}\right\} $ is the damping rate
matrix, $\mu =mn$, and $\mu ^{\prime }=m^{\prime }n^{\prime }$. In Eq. (\ref%
{o-1}) the Lamb shift matrix has been neglected. The matrix $\mathbf{L}$ is
a time-independent, real, and symmetric matrix. Its matrix elements are
given by Eqs. (\ref{m31}) and (\ref{m32}). The matrix $\mathbf{R}$ is also a
time-independent and real but non-symmetric matrix. Its matrix elements are
given by Eq. (\ref{3-3-1}).

To solve Eq. (\ref{o-1}), we use the split-operator method. \cite%
{Hermann1988} Using this method, the propagation of the density matrix is
calculated by%
\begin{equation}
\mathbf{\rho }\left( t+\Delta t\right) =\mathbf{P}_{L}\left( t\right) 
\mathbf{P}_{R}\left( t\right) \mathbf{P}_{L}\left( t\right) \mathbf{\rho }%
\left( t\right) ,  \label{o-2}
\end{equation}%
where, $\mathbf{P}_{L}\left( t\right) $ and $\mathbf{P}_{R}\left( t\right) $
are the propagators. They are given by%
\begin{equation}
\mathbf{P}_{\Theta }\left( t\right) =\exp \left( \lambda \mathbf{Q}\Delta
t\right) ,  \label{o-3}
\end{equation}%
where, $\lambda =-i/2$ and $\mathbf{Q}=\mathbf{L}\left( t+\Delta t/2\right) $
for $\Theta =L$, and $\lambda =1$ and $\mathbf{Q}=\mathbf{R}\left( t+\Delta
t/2\right) $ for $\Theta =R$.

Suppose that $\mathbf{Q}$ is an $N\times N$ matrix, and its eigenvalue and
right eigenvector are $q_{k}$ and $B_{k}=\left[ b_{1k},b_{2k},\cdots ,b_{Nk}%
\right] ^{T}$, respectively, where $T$ denotes the transpose of the vector.
We assume that $\mathbf{q}=\left\{ q_{k}\delta _{k^{\prime }k}\right\} $ and 
$\mathbf{B}=\left\{ b_{k^{\prime }k}\right\} $ are the two matrices
constructed respectively by the eigenvalue $q_{k}$ and eigenvector $B_{k}$,
then the matrix $\mathbf{Q}$ can be calculated by%
\begin{equation}
\mathbf{Q}=\mathbf{BqB}^{-1},  \label{o-5}
\end{equation}%
where, $\mathbf{B}^{-1}$ is the inverse matrix of $\mathbf{B}$. Substituting
Eq. (\ref{o-5}) into Eq. (\ref{o-3}), the propagator $\mathbf{P}_{\Theta
}\left( t\right) $ is then given by%
\begin{equation}
\mathbf{P}_{\Theta }\left( t\right) =\mathbf{B}\exp \left( \lambda \mathbf{q}%
\Delta t\right) \mathbf{B}^{-1},  \label{o-5-1}
\end{equation}%
where, $\exp \left( \lambda \mathbf{q}\Delta t\right) $ is a diagonal matrix
with non-zero diagonal matrix elements given by $\exp \left( \lambda
q_{k}\Delta t\right) $. If $\mathbf{Q}$ is a Hermitian operator $\mathbf{B}%
^{-1}=\mathbf{B}^{\dag }$ and if $\mathbf{Q}$ is a real and symmetric
operator (e.g., $\mathbf{L}$) $\mathbf{B}^{-1}=\mathbf{B}^{T}$.

\subsection{Effect of driving fields on relaxation and decoherence times}

By numerically solving the master equation in terms of the numerical method
introduced in the preceding section, population and coherence of the
microwave-driven 2D SQUID qubit are calculated, from which the relaxation
and decoherence times are extracted.

\subsubsection{Free decay of the 2D SQUID qubit}

To numerically calculate the relaxation and decoherence times of the 2D
SQUID qubit in the absence of driving fields (free decay) we assume that the
initial state of the qubit is a superposition state of the two computational
states $\left\vert 1\right\rangle $ and $\left\vert 2\right\rangle $ with $%
\rho _{11}\left( 0\right) =\rho _{12}\left( 0\right) =\rho _{21}\left(
0\right) =\rho _{22}\left( 0\right) =0.5$ and $\rho _{ij}(0)=0$ for all the
other combination of $i$ and $j$. To take into account the leakage to
non-computational states we include four levels $\left( N=4\right) $ in the
calculation. In the case of weak driving fields and weak damping, the
calculation with four levels $\left( N=4\right) $ is converged. By
numerically solving the master equation (\ref{o-1}) with the aforementioned
initial state we obtain the population and coherence of the 2D SQUID qubit
in free decay. Since the coherence is usually a complex and fast
oscillating, we use $\left\vert \rho _{12}\right\vert ^{2}$ instead of $\rho
_{12}$ to estimate the decoherence time. In FIG. 6 and FIG. 7, we show with
the solid lines the evolution of population inversion $\left( \rho
_{22}-\rho _{11}\right) $ and the squared modulus of coherence $\left\vert
\rho _{12}\right\vert ^{2}$, respectively.

For the DTLS, the population inversion and squared modulus of coherence in
free decay undergo simple exponential decays with \cite{zhou2006field_effect}
\begin{equation}
\rho _{22}-\rho _{11}=y_{1}+z_{1}e^{-t/\tau _{1}},  \label{f-1}
\end{equation}%
and%
\begin{equation}
\left\vert \rho _{12}\right\vert ^{2}=y_{2}+z_{2}e^{-2t/\tau _{2}},
\label{f-2}
\end{equation}%
where, the parameters $\tau _{1}$ and $\tau _{2}$ are the relaxation and
decoherence times of the DTLS in free decay, respectively.

To calculate the relaxation and decoherence times of the 2D SQUID qubit in
free decay, we fit the numerical results of $\left( \rho _{22}-\rho
_{11}\right) $ and $\left\vert \rho _{12}\right\vert ^{2}$ to the above
exponential functions. The results of least-square fitting are plotted in
FIG. 6 and FIG. 7 with the dashed lines, from which we obtain $T_{1}=\tau
_{1}=3.429$ $\mu $s and $T_{2}=\tau _{2}=2.243$ $\mu $s. The calculated
relaxation time is in very good agreement with the experimental result $%
T_{1}=3.45$ $\mu $s, \cite{Lishaoxiong06} demonstrating the validity of our
approach and calculation. Using Eq. (\ref{x-81-1}), we obtain the dephasing
time $T_{\varphi }=3.333$ $\mu $s. For comparison, we also calculate the
relaxation and decoherence times using the analytical expressions of the
DTLS in free decay given by Eqs. (\ref{x-81-1-3-1}) to (\ref{x-81-1-4}). The
results $T_{1}=3.429$ $\mu $s, $T_{2}=2.243$ $\mu $s, and $T_{\varphi }=3.333
$ $\mu $s are exactly same as the numerical results. In addition, since the
decoherence time is shorter than the relaxation time the dephasing is the
main source of decoherence.

\subsubsection{Rabi oscillation of the resonantly driven 2D SQUID qubit}

To calculate the relaxation and decoherence times of the 2D SQUID qubit in
the presence of a microwave field, we assume that the initial state of the
qubit is the ground state with $\rho _{11}(0)=1$ and $\rho _{ij}(0)=0$ for
all the other $i$ and $j$. By numerically solving the master equation (\ref%
{o-1}) with this initial condition, the population and coherence and thus
the relaxation and decoherence times of the driven 2D SQUID qubit are
calculated. The relaxation and decoherence times of the driven qubit depend
on the relative value of the field strength to the damping strength. \cite%
{zhou2006field_effect} In the underdamped regime for which the field
strength is larger than the damping strength, the decoherence can be
decomposed into intrinsic and field-induced ones. If the initial state of
the qubit is the ground state, the intrinsic decoherence vanishes and the
qubit has a single decoherence time which equals to the field-induced
decoherence time.

In FIG. 8 and FIG. 9 we plot with the solid lines the evolution of
population difference $\left( \rho _{11}-\rho _{22}\right) $ and squared
modulus of coherence $\left\vert \rho _{12}\right\vert ^{2}$, respectively,
for the 2D SQUID qubit resonantly driven by the microwave field with $\phi
_{\mu }=1.0\times 10^{-5}$ and $\omega _{\mu }=\omega _{21}=0.127\omega _{LC}
$. As shown in these figures, both $\left( \rho _{11}-\rho _{22}\right) $
and $\left\vert \rho _{12}\right\vert ^{2}$ undergo damped Rabi oscillations.

In the underdamped regime, the population difference and squared modulus of
coherence of the resonantly driven DTLS from an eigenstate undergo damped
Rabi oscillations as \cite{zhou2006field_effect} 
\begin{equation}
\rho _{11}-\rho _{22}=\widetilde{y}_{1}+\widetilde{z}_{1}\sin \left( \Omega
t+\varphi _{1}\right) e^{-t/\widetilde{\tau }_{1}},  \label{fit-3}
\end{equation}

and%
\begin{eqnarray}
\left\vert \rho _{12}\right\vert ^{2} &=&\widetilde{y}_{2}+\widetilde{z}%
_{2}\sin \left( \Omega t+\varphi _{2}\right) e^{-t/\widetilde{\tau }_{2}} 
\notag \\
&&+\widetilde{z}_{3}\sin ^{2}\left( \Omega t+\varphi _{2}\right) e^{-2t/%
\widetilde{\tau }_{2}},  \label{fit-4}
\end{eqnarray}%
where, $\Omega $ is the Rabi frequency, and $\widetilde{\tau }_{1}$ and $%
\widetilde{\tau }_{2}$ are the relaxation and (field-induced) decoherence
times of the driven DTLS, respectively.

To extract the relaxation and decoherence times of the driven 2D SQUID
qubit, we fit the calculated $\left( \rho _{11}-\rho _{22}\right) $ and $%
\left\vert \rho _{12}\right\vert ^{2}$ to the aforementioned exponentially
damped Rabi oscillating functions. The results of the best fit are shown in
FIG. 8 and FIG. 9 with dashed lines, from which we obtain $\Omega
=4.016\times 10^{-5}\omega _{LC}$, $\widetilde{T}_{1}=\widetilde{\tau }%
_{1}=2.689$ $\mu $s, and $\widetilde{T}_{22}=\widetilde{\tau }_{2}=2.682$ $%
\mu $s.

Using the same procedure, we have calculated the relaxation and decoherence
times of the 2D SQUID qubit resonantly driven by the microwave fields with
different field strengths. The results are given in the columns with $N=4$
in TABLE \ref{Table 2}. To examine the effect of leakage on the relaxation
and decoherence times, we have also calculated the relaxation and
decoherence times only using the two computational states. The results are
given in the columns with $N=2$ in TABLE \ref{Table 2}. In the case of weak
driving fields and weak damping, the 2D SQUID qubit may be well approximated
by a DTLS. Form the analytical expressions of relaxation and decoherence
times of the driven DTLS given by Eqs. (\ref{x-86}), (\ref{x-81-1-3-1}), and
(\ref{x-81-3}), we obtain $\widetilde{T}_{1}=\widetilde{T}_{22}=2.712$ $\mu $%
s. They are independent of driving field strength.

TABLE \ref{Table 2} shows that when $\phi _{\mu }\leq 1\times 10^{-6}$ the
calculated relaxation and decoherence times of the driven 2D SQUID qubit are
essentially identical. They are independent of the driving field strengths,
shorter than the relaxation time and longer than the decoherence time of the
2D SQUID qubit in free decay. These results accord with those obtained from
the calculation with the analytical expressions of characteristic times of
the driven DTLS. The relaxation and decoherence times obtained from the
calculation with $N=4$ are the same as those with $N=2$ and those with the
analytical expressions, demonstrating that both the strong field effect and
leakage are negligibly small in this case. When $\phi _{\mu }\geq 5\times
10^{-6}$, on one hand, the relaxation time obtained from the calculation
with $N=4$ equals to that with $N=2$, indicating that the leakage does not
influence the relaxation time. On the other hand, the relaxation time
obtained from the calculation with $N=2$ is less than that obtained from the
calculation with the analytical expressions, illustrating that the strong
field effect makes the relaxation time smaller. In contrast, due to the
strong field effect the decoherence time obtained from the calculation with $%
N=2$ increases slowly with the field strength, while due to the leakage the
decoherence time obtained from the calculation with $N=4$ decreases with the
field strength quickly. In particular, when $\phi _{\mu }\geq 1\times
10^{-4} $ the squared modulus of coherence obtained from the calculation
with $N=4$ no longer undergoes the simple damped Rabi oscillation as that
given by Eq. (\ref{fit-4}). Thus the relaxation time is sensitive to the
strong field effect while the decoherence time is sensitive to the leakage.

\begin{table}[htbp] \centering%
\caption{Numerical results of relaxation and decoherence times ($\mu$s) 
of the 2D SQUID qubit in free and driven decays.\label{Table 2}}%
\begin{tabular}{ccccccccc}
\hline\hline
Field strength & \ \  & Relaxation &  & time & \ \ \  & Decoherence &  & time
\\ \cline{3-5}\cline{7-9}
&  & $N=4$ &  & $N=2$ &  & $N=4$ &  & $N=2$ \\ \hline
$0$ &  & $3.429$ &  & $3.429$ &  & $2.243$ &  & $2.243$ \\ 
$1\times 10^{-7}$ &  & $2.712$ &  & $2.712$ &  & $2.712$ &  & $2.712$ \\ 
$5\times 10^{-7}$ &  & $2.712$ &  & $2.712$ &  & $2.712$ &  & $2.712$ \\ 
$1\times 10^{-6}$ &  & $2.712$ &  & $2.712$ &  & $2.712$ &  & $2.712$ \\ 
$5\times 10^{-6}$ &  & $2.706$ &  & $2.706$ &  & $2.705$ &  & $2.713$ \\ 
$1\times 10^{-5}$ &  & $2.689$ &  & $2.689$ &  & $2.682$ &  & $2.716$ \\ 
$5\times 10^{-5}$ &  & $2.224$ &  & $2.224$ &  & $1.945$ &  & $2.742$ \\ 
$1\times 10^{-4}$ &  & $1.480$ &  & $1.480$ &  & $-$ &  & $2.837$ \\ 
\hline\hline
\end{tabular}%
\end{table}%

\subsection{Optimization of the control and readout circuits}

The decoherence of the 2D SQUID qubit strongly depends on the control and
readout circuits. To optimize the control and readout circuits for long
decoherence time, we investigate how the characteristic times change with
the parameters and temperature of the circuits of the 2D SQUID qubit in free
and driven decays. The strength of the resonant microwave field used is $%
\phi _{\mu }=1\times 10^{-5}$.

\subsubsection{Characteristic times versus mutual inductances}

In FIG. 10 we show the characteristic times versus the mutual inductance $%
M_{x}$ between the 2D SQUID qubit and the control circuit. In this figure, $%
T_{1}$ and $T_{2}$ are the relaxation and decoherence times of the qubit in
free decay, and $\widetilde{T}_{1}$ and $\widetilde{T}_{22}$ are those of
the qubit in driven decay. It is shown that for all the values of $M_{x}$
shown in this figure $\widetilde{T}_{1}\simeq \widetilde{T}_{22}$ and $\min
\left( T_{1},T_{2}\right) \lesssim \widetilde{T}_{1}\lesssim \max \left(
T_{1},T_{2}\right) $. These results agree with the predictions from the
analytical expressions for the DTLS given by Eq. (\ref{x-86}) and hold, as
will be shown, for different control- and readout-circuit parameters. It is
also shown that when $M_{x}$ is \ less than $0.6$ pH $T_{2}>T_{1}$, when $%
M_{x}$ is larger than $0.6$ pH $T_{1}>T_{2}$, and when $M_{x}$ is equal to $%
0.6$ pH $T_{1}\simeq T_{2}$. Hence the relaxation is the dominating source
of decoherence for smaller $M_{x}$ while the dephasing is the main source of
decoherence for larger $M_{x}$. In addition, when $M_{x}<0.1$ pH the
characteristic times do not change with $M_{x}$ and when $M_{x}\geq 0.1$ pH
the characteristic times decrease with $M_{x}$ monotonically. These results
can be analyzed by using the analytical expressions of characteristic times
for the DTLS. For example, from Eqs. (\ref{x-81-1-3-1}), (\ref{i3-2-2}), (%
\ref{A-a10}), and (\ref{A-a11}), for small and large $M_{x}$, the relaxation
time $T_{1}$ of the qubit in free decay can be well approximated by 
\begin{equation}
T_{1}^{-1}\sim a_{0}\left( 1+b_{0}M_{x}^{2}\right) ,  \label{r-1}
\end{equation}%
where $a_{0}$ and $b_{0}$ are two parameters independent of $M_{x}$. Eq. (%
\ref{r-1}) shows that $T_{1}$ is a constant for $M_{x}\ll 1/\sqrt{b_{0}}$
and $T_{1}\propto M_{x}^{-2}$ for $M_{x}\gg 1/\sqrt{b_{0}}$, which are in
good agreement with the numerical results shown in FIG. 10. The changes of $%
T_{2}$, $\widetilde{T}_{1}$, and $\widetilde{T}_{22}$ with $M_{x}$ are very
similar to that of $T_{1}$ with $M_{x}$.

In FIG. 11 the characteristic times are shown versus the mutual inductance $%
M_{m}$ between the 2D SQUID qubit and the readout circuit. It is shown that
when $M_{m}$ is less than $6$ pH $T_{1}>T_{2}$ and the dephasing is the main
source of decoherence, when $M_{m}$ is larger than $6$ pH $T_{2}>T_{1}$ and
the relaxation is the dominating source of decoherence, and when $%
M_{m}\simeq 6$ pH $T_{1}\simeq T_{2}$. It is also shown that when $M_{m}<0.5$
pH the characteristic times do not change with $M_{m}$ and when $M_{m}\geq
0.5$ pH the characteristic times decrease monotonically with $M_{m}$. These
behaviors are very similar to those of the characteristic times versus $M_{x}
$. Thus the dependence of the characteristic times on $M_{m}$ can also be
well approximated by an equation analogous to Eq. (\ref{r-1}).

\subsubsection{Characteristic times versus readout-circuit parameters}

In FIG. 12 to FIG. 15, we plot the characteristic times versus the
inductances $L_{J1}$, $L_{J2}$, $L_{10}$, and $L_{20}$ of the readout
circuit, respectively. It is shown that the characteristic times change
dramatically with $L_{J1}$, $L_{J2}$, and $L_{10}$ but decrease slowly with $%
L_{20}$. When $L_{J1}$ $\left( L_{J2}\text{ or }L_{10}\right) $ is less than 
$550$ pH $\left( 100\text{ pH or }470\text{ pH}\right) $ the characteristic
times increase with $L_{J1}$ $\left( L_{J2}\text{ or }L_{10}\right) $. When $%
L_{J1}$ $\left( L_{J2}\text{ or }L_{10}\right) $ equals to $550$ pH $\left(
100\text{ pH or }470\text{ pH}\right) $ the characteristic times reach the
maxima. After $L_{J1}$ $\left( L_{J2}\text{ or }L_{10}\right) $\ is larger
than $550$ pH $\left( 100\text{ pH or }470\text{ pH}\right) $ the
characteristic times decrease with $L_{J1}$ $\left( L_{J2}\ \text{or }%
L_{10}\right) $. In contrast, the characteristic times do not change much
with $L_{20}$ for very small and very large $L_{20}$ and decrease with $%
L_{20}$ for moderately large $L_{20}$. In addition, for all the values of
inductances shown in these figures, $T_{1}>T_{2}$. Thus the dephasing is the
dominating source of decoherence

To gain insights into mechanisms behind these behaviors, we analyze these
results using the analytical expressions of characteristic times for the
DTLS. From Eqs. (\ref{x-81-1-3-1}), (\ref{i3-2-3}), and (\ref{x-9}), at the
adjacency of $\Delta L=0$ the relaxation time $T_{1}$ of the qubit in free
decay can be approximated by%
\begin{equation}
T_{1}^{-1}\sim a_{1}+b_{1}\left( \Delta L\right) ^{2},  \label{r-7}
\end{equation}%
where, $\Delta L=\left( L_{20}+L_{J2}\right) -\left( L_{10}+L_{J1}\right) $, 
$a_{1}$ is a parameter independent of $\Delta L$ and $b_{1}$ is a parameter
slowly and smoothly varying with $\Delta L$. Eq. (\ref{r-7}) shows that $%
T_{1}$ decreases with $\Delta L$ and reaches the maximum at $\Delta L=0$.

For the results in FIG. 12, the position of the maxima of the characteristic
times is at $L_{J1}=550$ pH. This result is in very good agreement with the
prediction from Eq. (\ref{r-7}) since when $L_{J1}=550$ pH $\Delta L=0$. For
the results in FIG. 13 and FIG. 14, we also have $\Delta L=0$ at the
positions of the maxima of the characteristic times. As for the results in
FIG. 15, $\Delta L$ is always larger than $430$ pH since $\Delta L=L_{20}+430
$ pH and $L_{20}\geq 0$. Thus the characteristic times do not have maxima on 
$L_{20}$.

In FIG. 16 we exhibit the characteristic times versus the capacitance $C_{m}$
of the readout circuit. When $C_{m}<0.01$ pF the characteristic times do not
change with $C_{m}$, when $0.01$ pF $\leq C_{m}\leq 2$ pF the characteristic
times decrease with $C_{m}$ , and when $C_{m}>2$ pF the characteristic times
tend to constants. For the values of $C_{m}$ shown in this figure, $%
T_{2}<T_{1}$ and thus the dephasing is the main source of decoherence.

In FIG. 17 and FIG. 18, we show the characteristic times versus the
resistances $R_{m}$ and $R_{m0}$ of the readout circuit, respectively. FIG.
17 shows that the characteristic times increase with $R_{m}$ when $%
R_{m}\lesssim 1000$ $\Omega $ and do not change with $R_{m}$ when $R_{m}>1000
$ $\Omega $. When $R_{m}$ is less than $21$ $\Omega $ $T_{2}>T_{1}$ and the
relaxation is the main source of decoherence, when $R_{m}$ is larger than $21
$ $\Omega $ $T_{1}>T_{2}$ and the dephasing is the dominating source of
decoherence, and when $R_{m}$ equals to $21$ $\Omega $ $T_{1}=T_{2}$. FIG.
18 shows that the characteristic times increase with $R_{m0}$ when $%
R_{m0}\lesssim 2000$ $\Omega $ and tend to constants after $R_{m0}>2000$ $%
\Omega $. For the values of $R_{m0}$ shown in the figure, $T_{1}>T_{2}$ and
the dephasing is the main source of decoherence.

\subsubsection{Characteristic times versus control-circuit parameters}

In FIG. 19 we plot the characteristic times versus the inductance $L_{x}$ of
the control circuit. It is shown that when $L_{x}\leq 1\times 10^{2}$ pH the
characteristic times are constants, when $1\times 10^{2}<L_{x}<1\times 10^{4}
$ pH the characteristic times increase with $L_{x}$, and when $L_{x}\geq
1\times 10^{4}$ pH the characteristic times do not change with $L_{x}$. For
all the values of $L_{x}$ shown in this figure $T_{2}<T_{1}$, demonstrating
that the dephasing is the main source of decoherence.

In FIG. 20 the characteristic times are plotted versus the capacitance $C_{x}
$ of the control circuit. It is shown that the changes of the characteristic
times with $C_{x}$ are very similar to those with $C_{m}$. When $C_{x}\leq
0.05$ pF the characteristic times do not change with $C_{x}$, when $%
0.05<C_{x}<2.5$ pF the characteristic times decrease with $C_{x}$, and when $%
C_{x}\geq 2.5$ pF the characteristic times tend to constants. For all the
values of $C_{x}$ shown in this figure $T_{2}<T_{1}$ and thus the dephasing
is the main source of decoherence.

In FIG. 21 and FIG. 22 the characteristic times are plotted versus the
resistances $R_{x}$ and $R_{x0}$ of the control circuit, respectively. It is
shown from FIG. 21 that when $R_{x}$ is less than $20$ $\Omega $ $T_{2}>T_{1}
$ and the relaxation is the main source of decoherence, when $R_{x}$ is
larger than $20$ $\Omega $ $T_{1}>T_{2}$ and the dephasing is the dominating
source of decoherence, and when $R_{x}$ equals to $20$ $\Omega $ $%
T_{1}\simeq T_{2}$. For $R_{x}<2000$ $\Omega $ the characteristic times
increase with $R_{x}$ and for $R_{x}\geq 2000$ $\Omega $ the characteristic
times tend to constants. The changes of the characteristic times with $R_{x0}
$ shown in FIG. 22 are very similar to those with $R_{x}$. When $R_{x0}$ is
less than $2200$ $\Omega $ $T_{1}>T_{2}$ and the dephasing is the main
source of decoherence, when $R_{x0}$ is larger than $2200$ $\Omega $ $%
T_{2}>T_{1}$ and the relaxation is the dominating source of decoherence, and
when $R_{x0}$ is equal to $2200$ $\Omega $ $T_{1}=T_{2}$. For $R_{x0}\leq
3\times 10^{4}$ $\Omega $ the characteristic times increase with $R_{x0}$
while for $R_{x0}>3\times 10^{4}$ $\Omega $ the characteristic times tend to
constants.

\subsubsection{Characteristic times versus temperature}

Finally, we plot the characteristic times versus the temperature $T$ of the
external circuits in FIG. 23. It is shown that in general the characteristic
times decrease with $T$. When $T\rightarrow 0$ the characteristic times tend
to constants and when $T\rightarrow \infty $ the characteristic times are
inversely proportional to $T$ which is the results of classical mechanics.
These results agree with the predictions from the analytical expressions of
characteristic times for the DTLS. It is also shown that at the lower
temperature when $T<10.3$ mK $T_{2}>T_{1}$ and the relaxation is the main
source of decoherence, while at the higher temperature when $T>10.3$ mK $%
T_{1}>T_{2}$ and the dephasing is the dominating source of decoherence.
These results also agree with the predictions from the analytical
expressions for the DTLS \cite{zhou2006field_effect} and with those obtained
by others. \cite{Leggett87,Tian2002,Burkard2004}.

\section{Conclusion}

In summary, to investigate the environment-induced decoherence in realistic
gate operations of solid-state qubits, we present a general theory for the
treatment of decoherence of a multilevel quantum system of many degrees of
freedom interacting with a multibath reservoir and driven by ac fields. In
this theory, the system is described by the reduced density operator
governed by the master equation. The effect of the environment on the system
is characterized by the spectral density through the dissipation
superoperator. The effects of driving field and leakage due to the coupling
with both the driving field and environment are included in this theory. In
the Hilbert space spanned by the eigenstates of the system's Hamiltonian,
the reduced density operator is represented by the density matrix and the
dissipation superoperator by the dissipation matrix. The diagonal and
off-diagonal matrix elements of the density matrix stand for the population
and coherence of the system, respectively. The dissipation matrix can be
decomposed into Lamb shift matrix and damping rate matrix. They are
determined by the transition matrix elements of the system and the spectral
density of the environment. The effect of the Lamb shift matrix on the
system is analogous to an extra field. In the case of weak damping the Lamb
shift matrix is extremely small compared to the driving field and are
neglected. In the study of decoherence of a qubit, for which the long-time
behavior of the qubit is significant, the damping rate matrix is replaced by
a steady one. For the thermal bath, the spontaneous decay rate and
stimulated transition rate are derived from the damping rate matrix. They
accord with those obtained by the others with different methods and obey the
detailed balance principle.

For an DTLS, the characteristic times in free and resonantly driven decays
are expressed by the analytical expressions. The decoherence of the driven
qubit can be decomposed into intrinsic and field-induced ones. The intrinsic
decoherence time equals to the decoherence time of the qubit in free decay
and the field-induced decoherence time equals to the relaxation time of the
driven qubit. In the case of weak driving fields, the relaxation and thus
the field-induced decoherence times of the driven qubit are independent of
the field strengths and they are always in between the relaxation and
decoherence times of the qubit in free decay.

For demonstration, we have applied the dissipative theory to simulate the
dissipation process of the 2D SQUID qubit coupled to the external circuits
in free decay. The energy levels, transition matrix elements, and relaxation
time are in very good agreement with the experimental results. We have also
applied the dissipative theory to investigate the effect of driving field
and leakage on the decoherence of the 2D SQUID qubit coupled to the external
circuits and resonantly driven by the microwave field. In the case of weak
driving fields, the relaxation and decoherence times of the driven qubit are
identical. They are independent of the driving field strength and in between
the relaxation and decoherence times of the qubit in free decay. These
results agree with the analytical results obtained from the analytical
expressions of characteristic times for the DTLS. In the case of a little
bit stronger driving fields, the relaxation time is sensitive to the strong
field effect while the decoherence time is sensitive to the leakage. In
addition, for the qubit in free decay, the relaxation is the main source of
decoherence at the low temperature while the dephasing is the dominating
source of decoherence at the high temperature.

To optimize the external circuits for long decoherence time, we have
investigated the characteristic times of the 2D SQUID qubit change with the
parameters and temperature of the control and readout circuits. We found
that the characteristic times decrease with the mutual inductances and
capacitances, increase with the resistances, and change dramatically with
the inductances, in particular, the inductances of the readout circuit. To
gain longer decoherence time, the coupling of the 2D SQUID qubit and the
external circuits should be weak, the capacitances of the external circuits
should be smaller, the resistances should be larger, the inductance of the
control circuit should be properly larger, and in particular, to reduce the
damping due to the readout circuit the total inductance of the left branch
of the readout circuit should balance with that of the right branch.

\begin{acknowledgments}
We acknowledge valuable discussions with Mr. Wei Qiu and Dr. Shaoxiong Li
about experimental setup. This work is supported in part by the NSF
(DMR-0325551) and by AFOSR, NSA, and ARDA through DURINT grant
(F49620-01-1-0439).
\end{acknowledgments}

\appendix

\section{Master equation of an open quantum system}

To derive the master equation for the reduced density operator of an open
quantum system, we work in interaction picture. In this picture the density
operator of the global system $\widetilde{\eta }(t)$ and the reduced density
operator of the quantum system $\widetilde{\rho }(t)$ are defined by%
\begin{equation}
\widetilde{\eta }(t)=\exp \left[ i\left( \mathcal{L}_{S}+\mathcal{L}%
_{R}\right) t\right] \eta (t),  \label{c7}
\end{equation}%
and%
\begin{equation}
\widetilde{\rho }(t)=\exp \left[ i\mathcal{L}_{S}t\right] \rho (t),
\label{a17-2}
\end{equation}%
where, $\eta (t)$ and $\rho (t)$ are the density operator of the global
system and the reduced density operator of the quantum system in the Schr%
\"{o}dinger picture, respectively. The relation between $\widetilde{\rho }%
(t) $ and $\widetilde{\eta }(t)$ is also given by Eq. (\ref{a17-1}).
Applying Eq. (\ref{c7}) to Eq. (\ref{a8}) we obtain the \textit{%
Liouville-von Neumann} for $\widetilde{\eta }(t)$ in the interaction picture 
\begin{equation}
\frac{d\widetilde{\eta }(t)}{dt}=-i\left[ \widetilde{\mathcal{L}}_{F}(t)+%
\widetilde{\mathcal{L}}_{I}(t)\right] \widetilde{\eta }(t),  \label{c8}
\end{equation}%
where, $\widetilde{\mathcal{L}}_{F}\left( t\right) $ and $\widetilde{%
\mathcal{L}}_{I}\left( t\right) $ are the presentations of $\mathcal{L}_{F}$
and $\mathcal{L}_{I}$ in the interaction picture. They are defined by%
\begin{equation}
\widetilde{\mathcal{L}}_{F}(t)=\exp \left( i\mathcal{L}_{S}t\right) \mathcal{%
L}_{F}\exp \left( -i\mathcal{L}_{S}t\right) ,  \label{a13}
\end{equation}%
and%
\begin{equation}
\widetilde{\mathcal{L}}_{I}(t)=\exp \left[ i\left( \mathcal{L}_{S}+\mathcal{L%
}_{R}\right) t\right] \mathcal{L}_{I}\exp \left[ -i\left( \mathcal{L}_{S}+%
\mathcal{L}_{R}\right) t\right] ,  \label{a14}
\end{equation}%
respectively.

The formal solution of Eq. (\ref{c8}) is%
\begin{equation}
\widetilde{\eta }(t)=\widehat{T}\exp \left[ -i\int_{0}^{t}\left( \widetilde{%
\mathcal{L}}_{F}(\tau )+\widetilde{\mathcal{L}}_{I}(\tau )\right) d\tau %
\right] \widetilde{\eta }(0),  \label{c10}
\end{equation}%
where, $\widehat{T}$ is the time-ordering operator and $\widetilde{\eta }(0)$
is the initial density operator of the global system which is equal to $\eta
(0)$ from Eq. (\ref{c7}).

Suppose the reservoir is uncorrelated with the system at $t=0$. In this
case, the initial density operator of the global system $\eta \left(
0\right) $ can be written as \cite{Gaspard1999}%
\begin{equation}
\eta \left( 0\right) =\rho \left( 0\right) \sigma \left( R\right) ,
\label{a11-01}
\end{equation}%
where, $\rho \left( 0\right) $ is the reduced density operator of the system
at $t=0$ and $\sigma \left( R\right) $ is the density operator of the
reservoir. For thermal baths at temperature $T$, for example, $\sigma \left(
R\right) $ is determined by the Boltzmann distribution \cite{Louisell1973}%
\begin{equation}
\sigma \left( R\right) =\frac{e^{-H_{R}/k_{B}T}}{\text{Tr}_{R}\left(
e^{-H_{R}/k_{B}T}\right) },  \label{a81}
\end{equation}%
where, $k_{B}$ is the Boltzmann constant. We also assume that (1) both the
interactions $H_{I}$ and $H_{F}$ are weak so that the Born perturbation
approximation can be applied to the exponential function of Eq. (\ref{c10}); 
\cite{Burkard2004} (2) the reservoir is sufficiently large and its states
are unperturbed by the coupling with the system and obey Gaussian statistics
so that $\sigma \left( R\right) $ is time-independent, Tr$_{R}\left[ \sigma
\left( R\right) \right] =1$, and Tr$_{R}\left[ H_{I}(t)\sigma \left(
R\right) \right] =0$; and (3) the characteristic time of correlation is much
less than the relaxation time of system so that the change of $\rho (t)$ is
slow and $\rho (\tau )$ can be replaced by $\rho (t)$ in the integral over
the correlation time. \cite{Kubo91} Under the these assumptions the system
will perform a Markovian process. \cite{Zoller2004} Substituting Eq. (\ref%
{c10}) into Eq. (\ref{a17-1}), calculating the trace over the reservoir by
means of the cumulant expansion method \cite{cumulant87} under the
aforementioned assumptions, and using the initial condition of Eq. (\ref%
{a11-01}), we finally obtain an equation for $\widetilde{\rho }(t)$ in a
series of cumulant, which is given, up to the second order of cumulant, by%
\begin{equation}
\widetilde{\rho }(t)=\widehat{T}\exp \left\{ -i\int_{0}^{t}\left[ \widetilde{%
\mathcal{L}}_{F}(\tau )+i\widetilde{\mathcal{D}}_{I}(\tau )\right] d\tau
\right\} \widetilde{\rho }(0),  \label{a22}
\end{equation}%
where, $\widetilde{\mathcal{D}}_{I}(t)$ is the dissipation superoperator in
the interaction picture given by%
\begin{equation}
\widetilde{\mathcal{D}}_{I}(t)=-\int_{0}^{t}\text{Tr}_{R}\left[ \widetilde{%
\mathcal{L}}_{I}\left( t\right) \widetilde{\mathcal{L}}_{I}\left( \tau
\right) \sigma \left( R\right) \right] d\tau .  \label{22-2}
\end{equation}

If making transformations $\widetilde{\eta }(t)\rightarrow \widetilde{\rho }%
(t)$ and $\widetilde{\mathcal{L}}_{I}(\tau )\rightarrow i\widetilde{\mathcal{%
D}}_{I}(\tau )$ in Eq. (\ref{c10}) one can get Eq. (\ref{a22}). Thus if
making the same transformations in Eq. (\ref{c8}) one obtains an equation of
motion for $\widetilde{\rho }(t)$%
\begin{equation}
\frac{d\widetilde{\rho }(t)}{dt}=-i\widetilde{\mathcal{L}}_{F}(t)\widetilde{%
\rho }(t)+\widetilde{\mathcal{D}}_{I}(t)\widetilde{\rho }(t),  \label{22-1}
\end{equation}%
This equation is recognized the master equation in the interaction picture.

Applying Eq. (\ref{a17-2}) to Eq. (\ref{22-1}), one obtains the master
equation for $\rho (t)$ in the Schr\"{o}dinger picture%
\begin{equation}
\frac{d\rho (t)}{dt}=-i\left[ \mathcal{L}_{S}+\mathcal{L}_{F}(t)\right] \rho
(t)+\mathcal{D}_{I}(t)\rho (t),  \label{22-1-1}
\end{equation}%
where, $\mathcal{D}_{I}\left( t\right) $ is the representation of $%
\widetilde{\mathcal{D}}_{I}(t)$ in the Schr\"{o}dinger picture given by%
\begin{equation}
\mathcal{D}_{I}(t)=-\int_{0}^{t}\text{Tr}_{R}\left[ \mathcal{L}%
_{I}^{S}\left( t,\tau \right) \mathcal{L}_{I}^{S}\left( \tau ,t\right)
\sigma \left( R\right) \right] d\tau ,  \label{a23}
\end{equation}%
with 
\begin{equation}
\mathcal{L}_{I}^{S}\left( t_{1},t_{2}\right) =\exp \left[ -i\mathcal{L}%
_{S}t_{1}\right] \widetilde{\mathcal{L}}_{I}\left( t_{1}\right) \exp \left[ i%
\mathcal{L}_{S}t_{2}\right] .  \label{a23-2}
\end{equation}

\section{Quantum Langevin equation and fluctuation-dissipation theorem}

We consider a quantum particle of mass $M$ moving in a potential $V\left(
q\right) $ and linearly coupled to a thermal bath at temperature $T$. In
phase space, the motion of the quantum particle is described by the quantum
Langevin equation \cite{Weiss1999}%
\begin{equation}
M\frac{d^{2}q(t)}{dt^{2}}+M\int_{t_{0}}^{t}d\tau \gamma \left( t-\tau
\right) \frac{dq(\tau )}{d\tau }+\frac{dV\left( q\right) }{dq}=\xi \left(
t\right) ,  \label{A1}
\end{equation}%
where, $q(t)$ is the coordinate operator of the particle, $\gamma \left(
t\right) $ is the damping coefficient, and $\xi \left( t\right) $ is the
fluctuation force of the thermal bath. The quantum Langevin equation
provides a relation between the damping coefficient $\gamma \left( t\right) $
and the fluctuation force $\xi \left( t\right) $ of the thermal bath.

In frequency domain, the frequency-dependent damping coefficient $\gamma
\left( \omega \right) $ is calculated from $\gamma \left( t\right) $ by
Fourier transform%
\begin{eqnarray}
\gamma \left( \omega \right) &=&\gamma _{R}\left( \omega \right) +i\gamma
_{I}\left( \omega \right)  \notag \\
&=&\int_{-\infty }^{+\infty }\gamma \left( t\right) \exp \left( -i\omega
t\right) dt.  \label{A2}
\end{eqnarray}%
The spectral density of the fluctuation force $\xi \left( t\right) $ at
temperature $T$, $J_{\xi }\left( \omega \right) $, can be calculated from $%
\gamma _{R}\left( \omega \right) $ in terms of the quantum
fluctuation-dissipation theorem \cite{Weiss1999,Devoret97} 
\begin{equation}
J_{\xi }\left( \omega \right) =M\hbar \omega \gamma _{R}\left( \omega
\right) \left[ 1+\coth \left( \frac{\hbar \omega }{2k_{B}T}\right) \right] .
\label{A3}
\end{equation}%
At high temperature limit, $T\gg \hbar \omega /k_{B}$ and $J_{\xi }\left(
\omega \right) $ in Eq. (\ref{A3}) is simplified to the result of the
quasiclassical and classical fluctuation-dissipation theorem \cite{Weiss1999}%
\begin{eqnarray}
J_{\xi }\left( \omega \right) &=&M\hbar \omega \gamma _{R}\left( \omega
\right) \coth \left( \frac{\hbar \omega }{2k_{B}T}\right)  \notag \\
&\simeq &2Mk_{B}T\gamma _{R}\left( \omega \right) .  \label{A4}
\end{eqnarray}%
It is shown that for a classical particle $J_{\xi }\left( \omega \right) $
is proportional to the temperature. In contrast, at low temperature limit, $%
T\ll \hbar \omega /k_{B}$ and $J_{\xi }\left( \omega \right) $ in Eq. (\ref%
{A3}) is simplified to%
\begin{equation}
J_{\xi }\left( \omega \right) =M\hbar \omega \gamma _{R}\left( \omega
\right) ,  \label{A4-1}
\end{equation}%
which is independent of the temperature.

The autocorrelation function of the fluctuation force $\xi \left( t\right) $%
, $\mathcal{J}_{\xi }\left( t\right) =\left\langle \xi \left( t\right) \xi
\left( 0\right) \right\rangle $, is calculated from $J_{\xi }\left( \omega
\right) $ by using Eq. (\ref{m26-3}). If $\gamma _{R}\left( \omega \right) $
is an even function of $\omega $, the autocorrelation function of a quantum
particle is given by \cite{Weiss1999}%
\begin{eqnarray}
\mathcal{J}_{\xi }\left( t\right) &=&\frac{\hbar M}{\pi }\int_{0}^{+\infty
}\omega \gamma _{R}\left( \omega \right) \left[ i\sin \left( \omega t\right)
\right.  \notag \\
&&+\coth \left( \frac{\hbar \omega }{2k_{B}T}\right) \left. \cos \left(
\omega t\right) \right] d\omega .  \label{A6}
\end{eqnarray}%
And the autocorrelation function of a quasiclassical particle is given by%
\begin{eqnarray}
\mathcal{J}_{\xi }\left( t\right) &=&\frac{\hbar M}{\pi }\int_{0}^{+\infty
}\omega \gamma _{R}\left( \omega \right) \coth \left( \frac{\hbar \omega }{%
2k_{B}T}\right)  \notag \\
&&\times \cos \left( \omega t\right) d\omega .  \label{A7}
\end{eqnarray}%
From Eq. (\ref{A7}) the autocorrelation function of the quasiclassical
particle is real and reversible, i.e., $\left\langle \xi \left( t\right) \xi
\left( 0\right) \right\rangle =\left\langle \xi \left( -t\right) \xi \left(
0\right) \right\rangle $. However, from Eq. (\ref{A6}) the autocorrelation
function of the quantum particle is complex and irreversible.

\section{Spectral density of the control and readout circuits}

For the 2D SQUID flux qubit shown in FIG. 1, the thermal bath consists of
the control and readout circuits. They produce thermal noises for the 2D
SQUID qubit.

\subsection{Spectral density of the control circuit}

As shown in FIG. 24, the control circuit is coupled to the rf SQUID through
the loop $L$ with mutual inductance $M_{x}$. The equivalent impedance $Z_{1}$
of the parallel branches $C_{x}R_{x}$ and $R_{x0}$ is given by%
\begin{equation}
Z_{1}=R_{x}^{\text{eq}}\left( \omega \right) +\frac{1}{j\omega C_{x}^{\text{%
eq}}\left( \omega \right) },  \label{z-2}
\end{equation}%
where, $R_{x}^{\text{eq}}\left( \omega \right) $ is the frequency-dependent
equivalent resistance given by%
\begin{equation}
R_{x}^{\text{eq}}\left( \omega \right) =\frac{1+\omega
^{2}C_{x}^{2}R_{x}\left( R_{x}+R_{x0}\right) }{1+\omega ^{2}C_{x}^{2}\left(
R_{x}+R_{x0}\right) ^{2}}R_{x0},  \label{z-3}
\end{equation}%
and $C_{x}^{\text{eq}}\left( \omega \right) $ is the frequency-dependent
equivalent capacitance given by%
\begin{equation}
C_{x}^{\text{eq}}\left( \omega \right) =\frac{1+\omega ^{2}C_{x}^{2}\left(
R_{x}+R_{x0}\right) ^{2}}{\omega ^{2}C_{x}\allowbreak R_{x0}^{2}}.
\label{z-4}
\end{equation}%
When $R_{x0}\rightarrow \infty $, $R_{x}^{\text{eq}}=R_{x}$ and $C_{x}^{%
\text{eq}}=C_{x}$. When $\omega \rightarrow 0$, $R_{x}^{\text{eq}}=R_{x0}$
and $\omega ^{2}C_{x}^{\text{eq}}=1/C_{x}\allowbreak R_{x0}^{2}$.

The circuit equations of the control circuit are%
\begin{equation}
j\omega LI-j\omega M_{x}I_{0}=U,  \label{A-a7}
\end{equation}%
\begin{equation}
-j\omega M_{x}I+\left( j\omega L_{x}+Z_{1}\right) I_{0}=0.  \label{A-a7-0}
\end{equation}%
From these equations, the equivalent impedance $Z_{x}\left( \omega \right) $
of the control circuit is given by%
\begin{equation}
Z_{x}\left( \omega \right) =\frac{U}{I}=j\omega L+\frac{\omega ^{2}M_{x}^{2}%
}{j\omega L_{x}+Z_{1}}.  \label{A-a8}
\end{equation}%
From this equation, the frequency-dependent equivalent admittance $%
Y_{x}(\omega )=1/Z_{x}(\omega )$ is obtained and its real part $Y_{xR}\left(
\omega \right) $ is given by%
\begin{equation}
Y_{xR}\left( \omega \right) =\frac{F_{x}\left( \omega \right) }{\omega
^{2}+G_{x}\left( \omega \right) },  \label{A-a9}
\end{equation}%
where,%
\begin{equation}
F_{x}\left( \omega \right) =\frac{M_{x}^{2}R_{x}^{\text{eq}}}{\upsilon
_{x}^{2}},  \label{A-a10}
\end{equation}%
\begin{equation}
G_{x}\left( \omega \right) =\frac{2L}{\upsilon _{x}C_{x}^{\text{eq}}}+\frac{%
L^{2}}{\upsilon _{x}^{2}}\left( R_{x}^{\text{eq}2}+\frac{1}{\omega
^{2}C_{x}^{\text{eq}2}}\right) ,  \label{A-a11}
\end{equation}%
and%
\begin{equation}
\upsilon _{x}=M_{x}^{2}-LL_{x}.  \label{A-a12}
\end{equation}

When $R_{x0}\rightarrow \infty $%
\begin{equation}
F_{x}\left( \omega \right) =\frac{M_{x}^{2}R_{x}}{\upsilon _{x}^{2}},
\label{n-1}
\end{equation}%
and%
\begin{equation}
G_{x}\left( \omega \right) =\frac{2L}{\upsilon _{x}C_{x}}+\frac{L^{2}}{%
\upsilon _{x}^{2}}\left( R_{x}^{2}+\frac{1}{\omega ^{2}C_{x}^{2}}\right) .
\label{n-2}
\end{equation}

When $\omega \rightarrow 0$%
\begin{equation}
F_{x}\left( \omega \right) =\frac{M_{x}^{2}R_{x0}}{\upsilon _{x}^{2}},
\label{n-3}
\end{equation}%
and%
\begin{equation}
G_{x}\left( \omega \right) =\frac{L^{2}R_{x0}^{2}}{\upsilon _{x}^{2}}.
\label{n-4}
\end{equation}

\subsection{\textbf{Spectral density of the readout circuit}}

As shown in FIG. 25, the readout circuit is coupled to the rf SQUID through
the loop $L$ with mutual inductance $M_{m}$. The equivalent impedance $Z_{2}$
of the parallel branches $C_{m}R_{m}$ and $R_{m0}$ is given by%
\begin{equation}
Z_{2}=R_{m}^{\text{eq}}\left( \omega \right) +\frac{1}{j\omega C_{m}^{\text{%
eq}}\left( \omega \right) },  \label{y-2}
\end{equation}%
where, $R_{m}^{\text{eq}}\left( \omega \right) $ is the frequency-dependent
equivalent resistance given by 
\begin{equation}
R_{m}^{\text{eq}}\left( \omega \right) =\frac{1+\omega
^{2}C_{m}^{2}R_{m}\left( R_{m}+R_{m0}\right) }{1+\omega ^{2}C_{m}^{2}\left(
R_{m}+R_{m0}\right) ^{2}}R_{m0},  \label{y-3}
\end{equation}%
and $C_{m}^{\text{eq}}\left( \omega \right) $ is the frequency-dependent
equivalent capacitance given by%
\begin{equation}
C_{m}^{\text{eq}}\left( \omega \right) =\frac{1+\omega ^{2}C_{m}^{2}\left(
R_{m}+R_{m0}\right) ^{2}}{\omega ^{2}C_{m}\allowbreak R_{m0}^{2}},
\label{y-4}
\end{equation}%
When $R_{m0}\rightarrow \infty $, $R_{m}^{\text{eq}}=R_{m}$ and $C_{m}^{%
\text{eq}}=C_{m}$. When $\omega \rightarrow 0$, $R_{m}^{\text{eq}}=R_{m0}$
and $\omega ^{2}C_{m}^{\text{eq}}=1/C_{m}\allowbreak R_{m0}^{2}$.

The circuit equations of the readout circuit are given by%
\begin{equation}
j\omega LI+j\omega \frac{M_{m}}{2}I_{1}-j\omega \frac{M_{m}}{2}I_{2}=U,
\label{1}
\end{equation}%
\begin{equation}
j\omega L_{1}I_{1}+j\omega \frac{M_{m}}{2}I=j\omega L_{2}I_{2}-j\omega \frac{%
M_{m}}{2}I,  \label{1-1}
\end{equation}%
and%
\begin{equation}
j\omega L_{1}I_{1}+j\omega \frac{M_{m}}{2}I=-\left( I_{1}+I_{2}\right) Z_{2},
\label{1-2}
\end{equation}%
where, $L_{1}=L_{10}+L_{J1}$ and $L_{2}=L_{20}+L_{J2}$. From these equations
the equivalent admittance of the readout circuit $Y_{m}\left( \omega \right)
=I/U$ can be obtained. Its real part is given by%
\begin{equation}
Y_{mR}\left( \omega \right) =\frac{1}{F_{m}\left( \omega \right) \left[
1+G_{m}\left( \omega \right) \omega ^{2}\right] },  \label{x-8}
\end{equation}%
where, $F_{m}\left( \omega \right) $ and $G_{m}\left( \omega \right) $ are
given by%
\begin{equation}
F_{m}\left( \omega \right) =R_{m}^{\text{eq}}\left( \dfrac{L}{M_{m}}\right)
^{2}\left( \dfrac{2L_{dc}}{\Delta L}\right) ^{2}\left( 1-k_{dc}^{2}\right)
^{2},  \label{x-9}
\end{equation}%
and%
\begin{equation}
G_{m}\left( \omega \right) =\frac{1}{R_{m}^{\text{eq2}}}\left[ \dfrac{%
L_{\parallel }\left( 1-k_{\parallel }^{2}\right) }{1-k_{dc}^{2}}-\frac{1}{%
\omega ^{2}C_{m}^{\text{eq}}}\right] ^{2},  \label{x-10}
\end{equation}%
where, $\Delta L=L_{2}-L_{1}$,\ $L_{dc}=L_{1}+L_{2}$, $L_{\parallel
}=L_{1}L_{2}/L_{dc}$, $k_{dc}^{2}=M_{m}^{2}/LL_{dc}$, and $k_{\parallel
}^{2}=M_{m}^{2}/4LL_{\parallel }$.

When $R_{m0}\rightarrow \infty $%
\begin{equation}
F_{m}=R_{m}\left( \dfrac{L}{M_{m}}\right) ^{2}\left( \dfrac{2L_{dc}}{\Delta L%
}\right) ^{2}\left( 1-k_{dc}^{2}\right) ^{2},  \label{x-11}
\end{equation}%
and%
\begin{equation}
G_{m}=\frac{1}{R_{m}^{2}}\left[ \dfrac{L_{\parallel }\left( 1-k_{\parallel
}^{2}\right) }{1-k_{dc}^{2}}-\frac{1}{\omega ^{2}C_{m}}\right] ^{2}.
\label{x-12}
\end{equation}%
When $\omega \rightarrow 0$%
\begin{equation}
F_{m}=R_{m0}\left( \dfrac{L}{M_{m}}\right) ^{2}\left( \dfrac{2L_{dc}}{\Delta
L}\right) ^{2}\left( 1-k_{dc}^{2}\right) ^{2},  \label{x-13}
\end{equation}%
and%
\begin{equation}
G_{m}=\frac{1}{R_{m0}^{2}}\left[ \dfrac{L_{\parallel }\left( 1-k_{\parallel
}^{2}\right) }{1-k_{dc}^{2}}-C_{m}\allowbreak R_{m0}^{2}\right] ^{2}.
\label{x-14}
\end{equation}%
\bibliographystyle{apsrev}
\bibliography{noise1}

\newpage 

Figure Captions

\bigskip 

FIG. 1 Sketch of the 2D SQUID qubit inductively coupled to the control and
readout circuits.

FIG. 2 (Online color) Contour of the potential energy of the 2D SQUID qubit.

FIG. 3 (Online color) Energy levels of the 2D SQUID qubit versus $x_{e}$.

FIG. 4 (Online color) Transition matrix elements of the 2D SQUID qubit
versus $x_{e}$.

FIG. 5 (Online color) $J_{x}(\omega )$, $J_{m}(\omega )$, and $J(\omega )$
versus $\omega $\ for the external circuits of the 2D SQUID flux qubit at $%
T=30$ mK.

FIG. 6 (Online color) Evolution of population inversion of the 2D SQUID
qubit in free decay. The solid and dashed lines are the numerical and
fitting results, respectively.

FIG. 7 (Online color) Same as FIG. 6 but for the squared modulus of
coherence.

FIG. 8 (Online color) Evolution of population difference of the 2D SQUID
qubit resonantly driven by the microwave field with $\phi _{\mu }=1.0\times
10^{-5}$ and $\omega _{\mu }=\omega _{21}=0.127\omega _{LC}$. The solid and
dashed lines are the numerical and fitting results, respectively.

FIG. 9 (Online color) Same as FIG. 8 but for the squared modulus of
coherence.

FIG. 10 (Online color) Characteristic times versus the mutual inductance $%
M_{x}$ between the 2D SQUID qubit and the control circuit. In this figure, $%
T_{1}$ and $T_{2}$ are the relaxation and decoherence times of the qubit in
free decay, and $\widetilde{T}_{1}$ and $\widetilde{T}_{22}$ are the
relaxation and decoherence times of the qubit in driven decay.

FIG. 11 (Online color) Same as FIG. 10 but for characteristic times versus
the mutual inductance $M_{m}$ between the 2D SQUID qubit and the readout
circuit.

FIG. 12 (Online color) Same as FIG. 10 but for characteristic times versus
the inductance $L_{J1}$ of the first junction of the readout circuit.

FIG. 13 (Online color) Same as FIG. 10 but for characteristic times versus
the inductance $L_{J2}$ of the second junction of the readout circuit.

FIG. 14 (Online color) Same as FIG. 10 but for characteristic times versus
the inductance $L_{10}$ of the readout circuit.

FIG. 15 (Online color) Same as FIG. 10 but for characteristic times versus
the inductance $L_{20}$ of the readout circuit.

FIG. 16 (Online color) Same as FIG. 10 but for characteristic times versus
the capacitance $C_{m}$ of the readout circuit.

FIG. 17 (Online color) Same as FIG. 10 but for characteristic times versus
the resistance $R_{m}$ of the readout circuit.

FIG. 18 (Online color) Same as FIG. 10 but for characteristic times versus
the resistance $R_{m0}$ of the readout circuit.

FIG. 19 (Online color) Same as FIG. 10 but for characteristic times versus
the inductance $L_{x}$ of the control circuit.

FIG. 20 (Online color) Same as FIG. 10 but for characteristic times versus
the capacitance $C_{x}$ of the control circuit.

FIG. 21 (Online color) Same as FIG. 10 but for characteristic times versus
the resistance $R_{x}$ of the control circuit.

FIG. 22 (Online color) Same as FIG. 10 but for characteristic times versus
the resistance $R_{x0}$ of the control circuit.

FIG. 23 (Online color) Same as FIG. 10 but for characteristic times versus
the temperature $T$.

FIG. 24 The control circuit of the 2D SQUID flux qubit and its equivalent
admittance $Y_{x}(\omega )$.

FIG. 25 The readout circuit of the 2D SQUID flux qubit and its equivalent
admittance $Y_{m}(\omega )$.

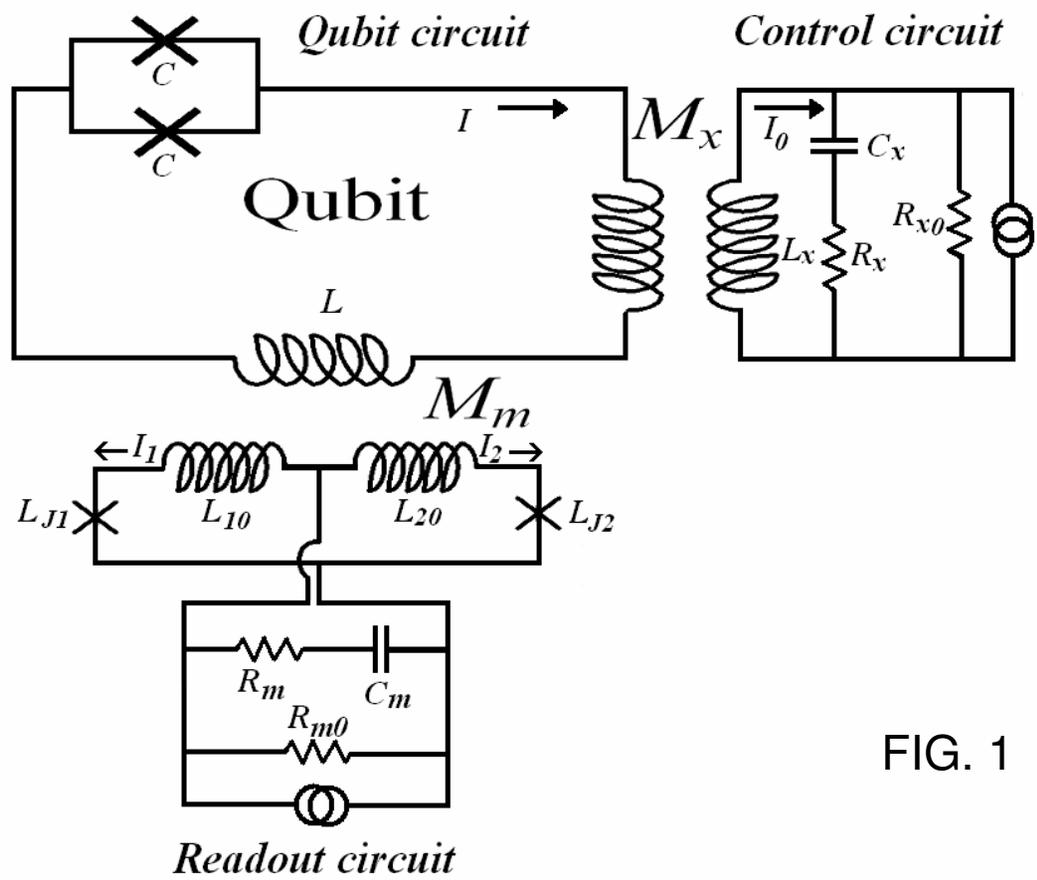
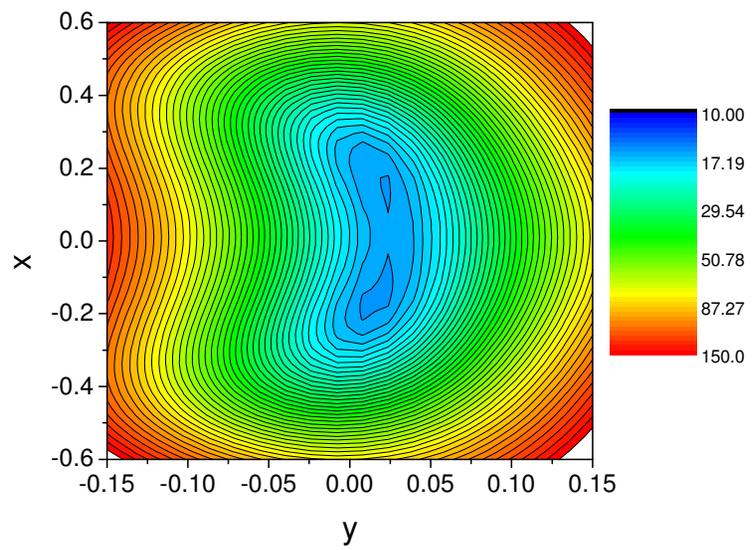
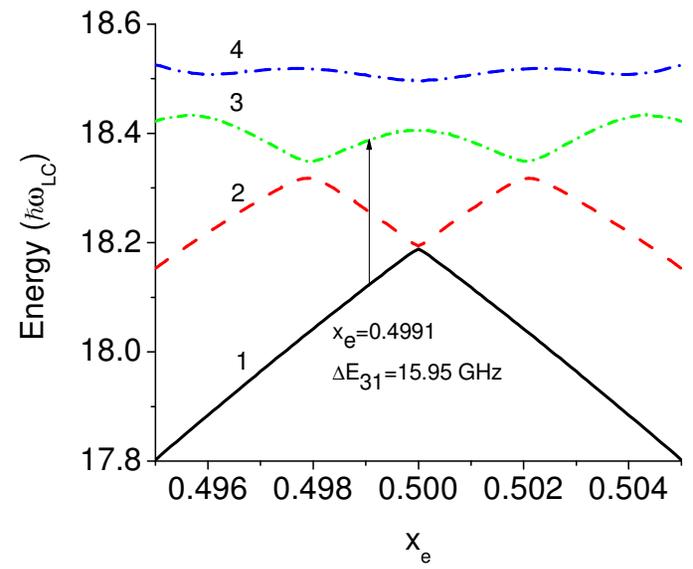
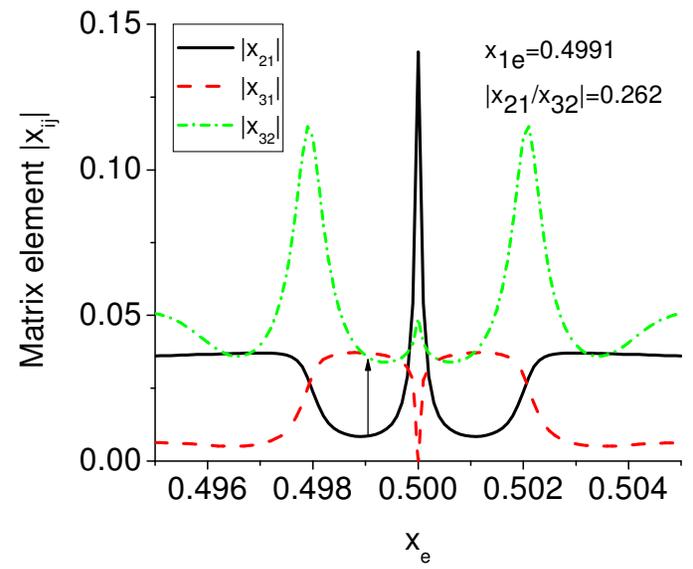
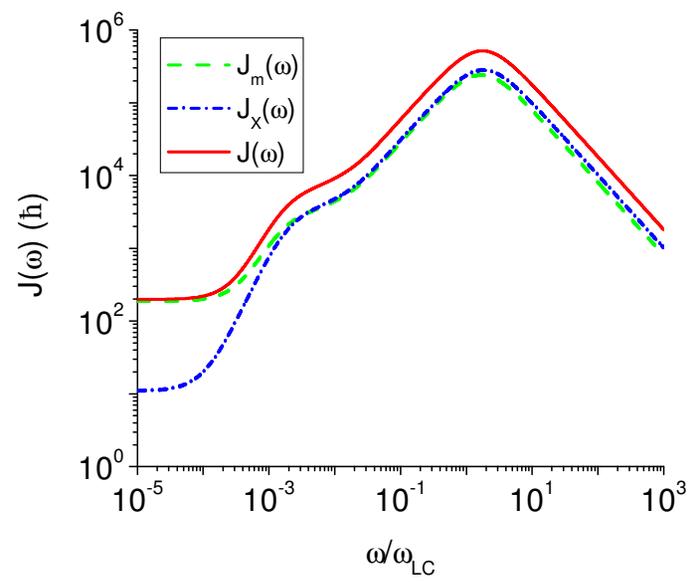
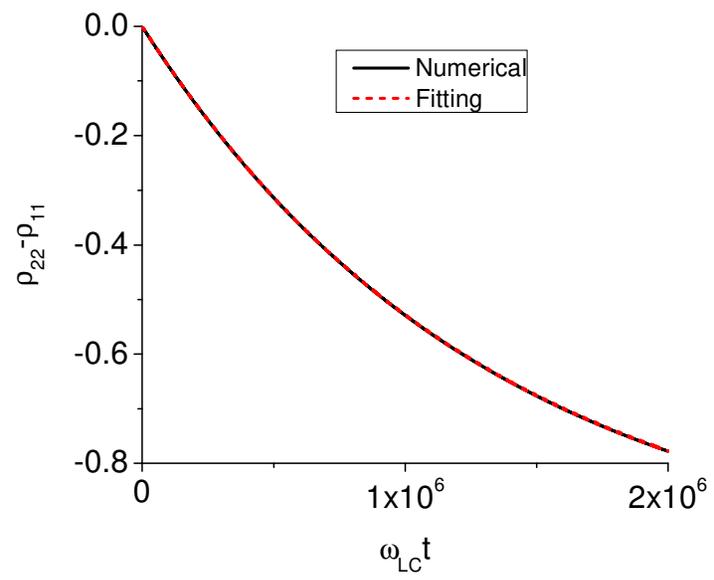
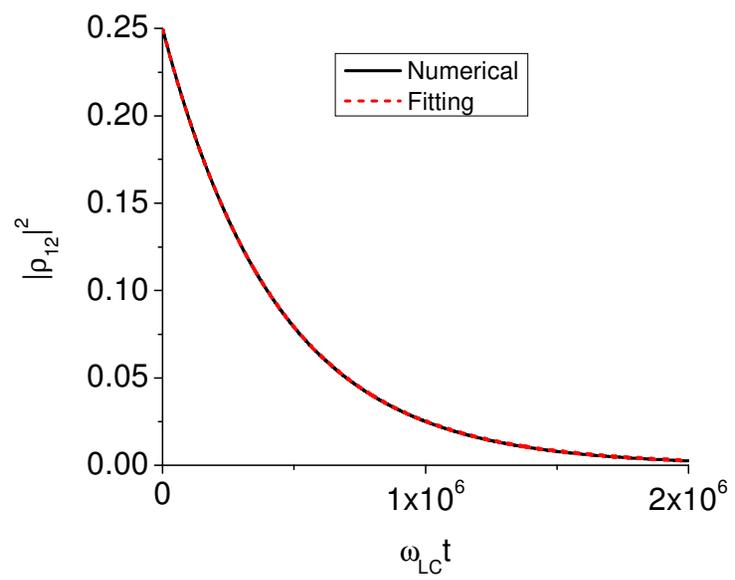
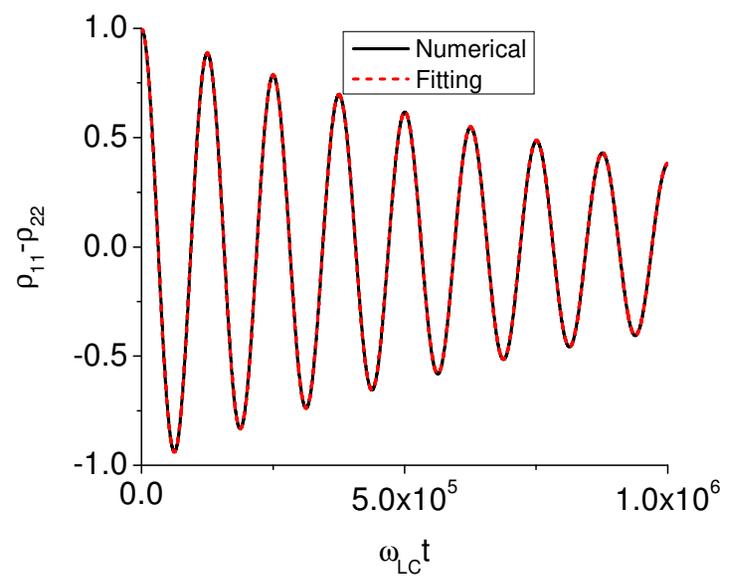
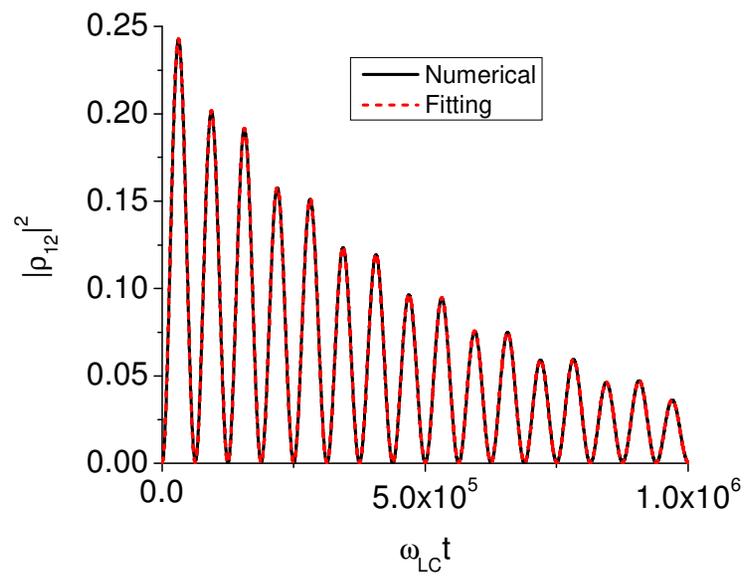
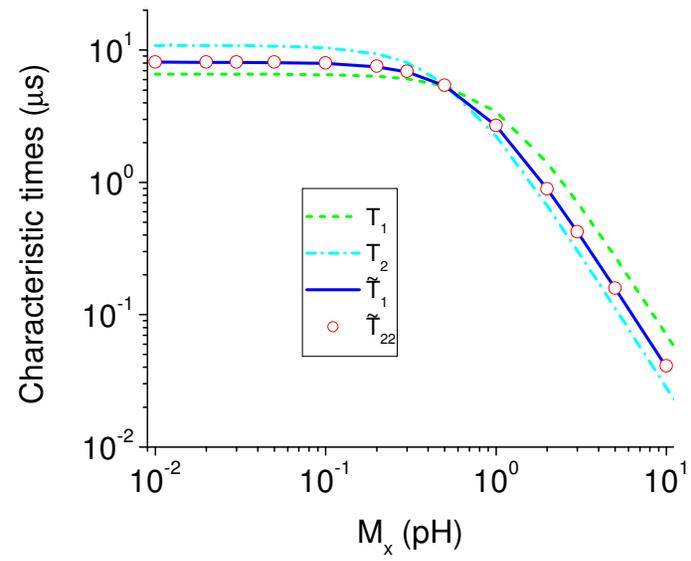
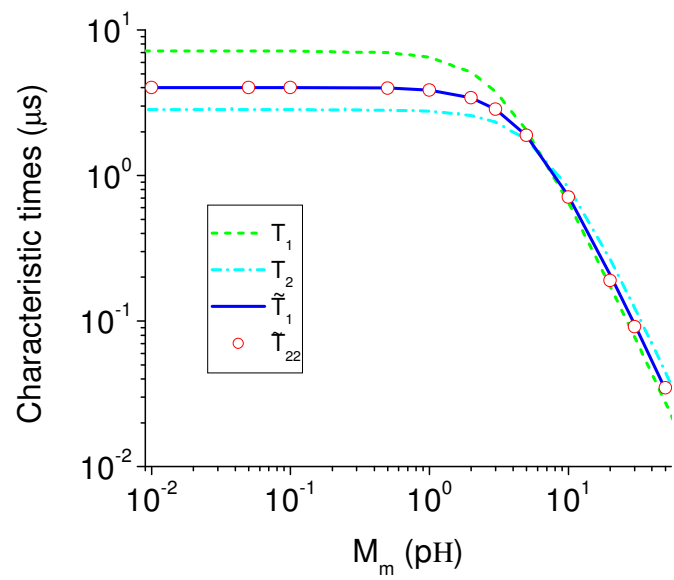
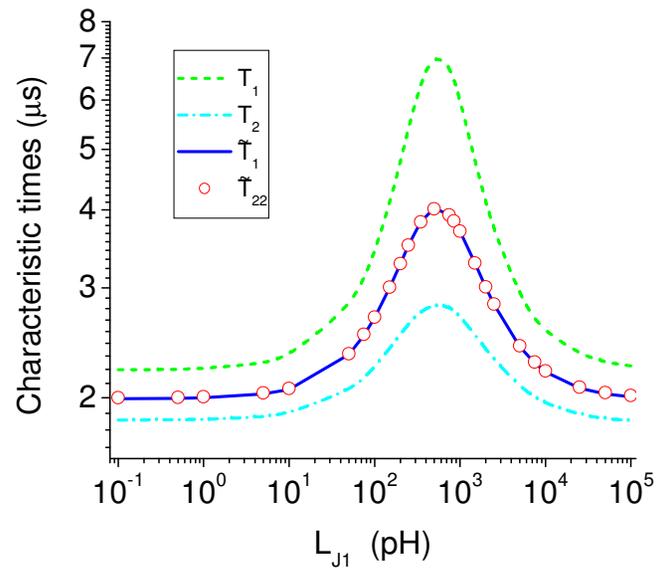
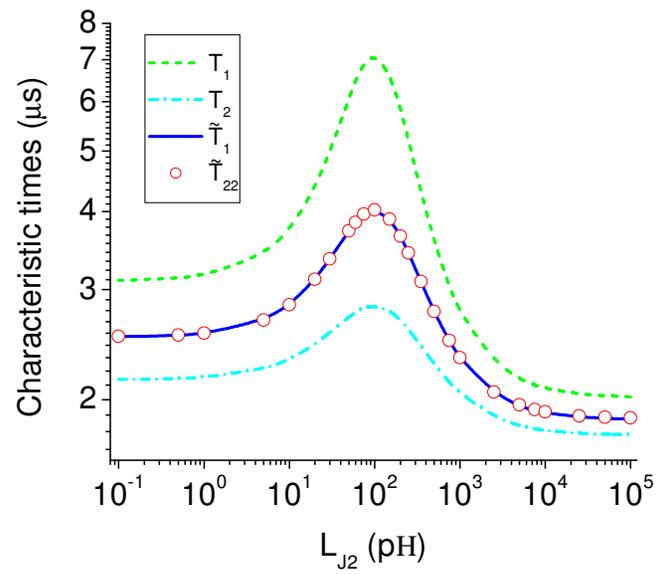
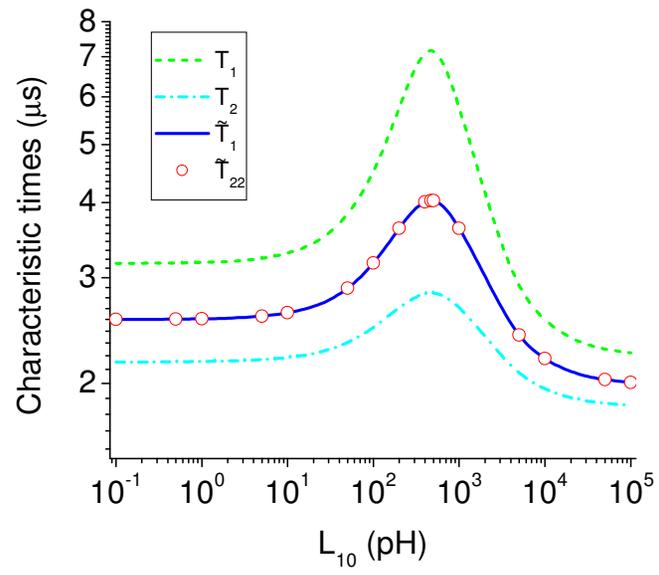
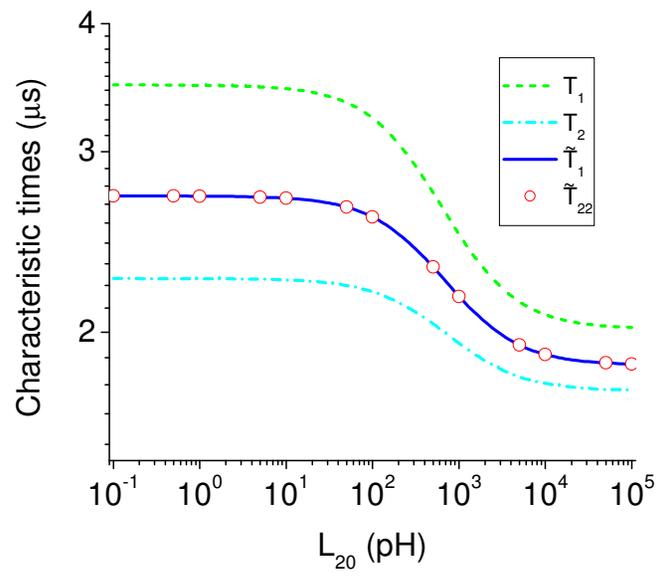
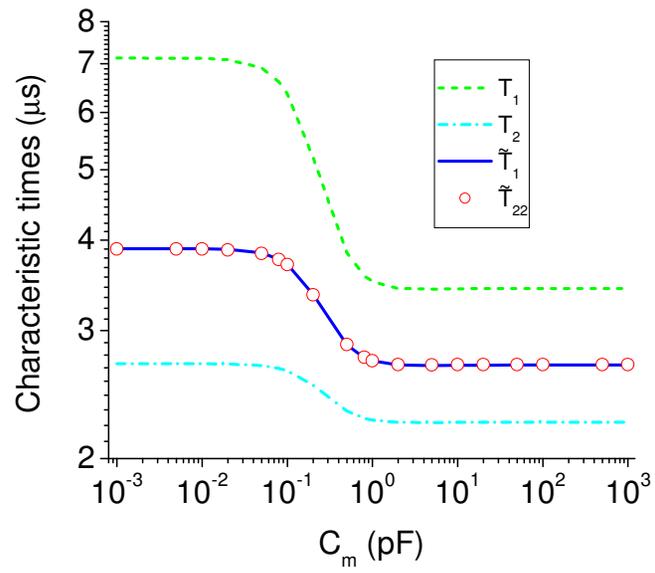
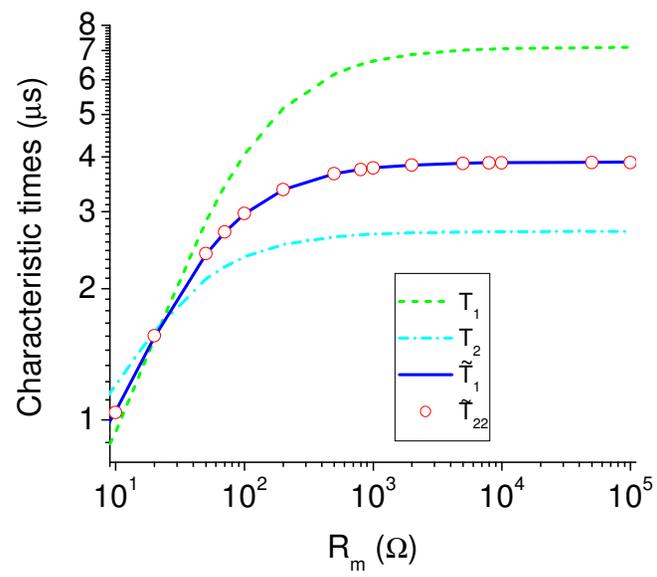
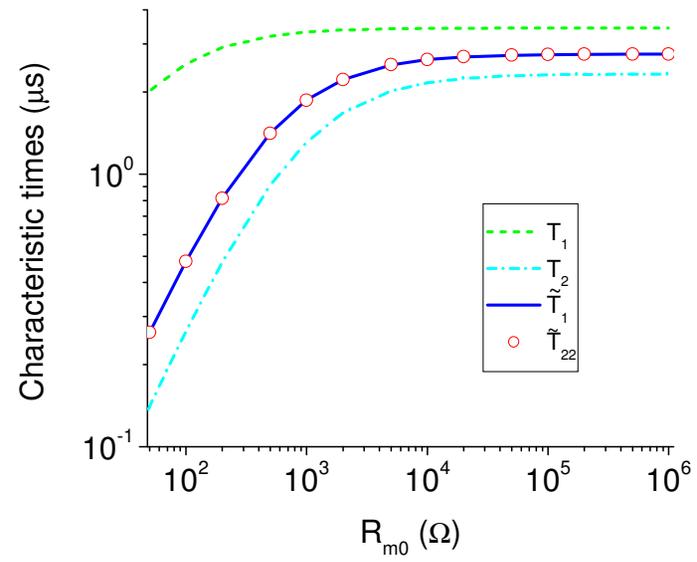
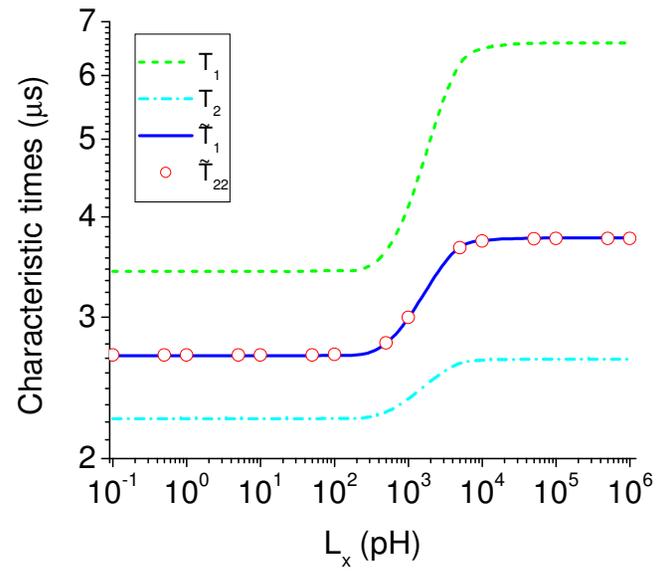
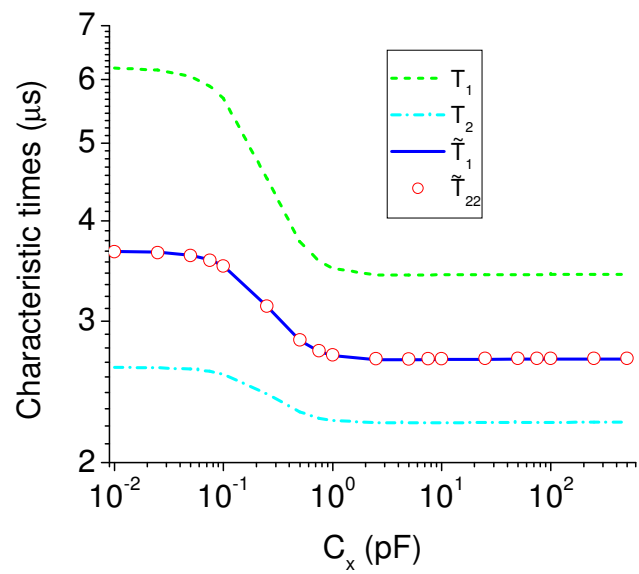
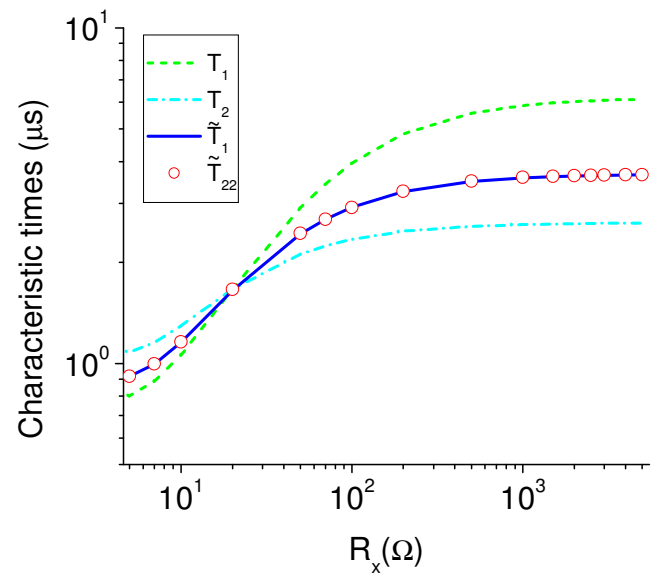
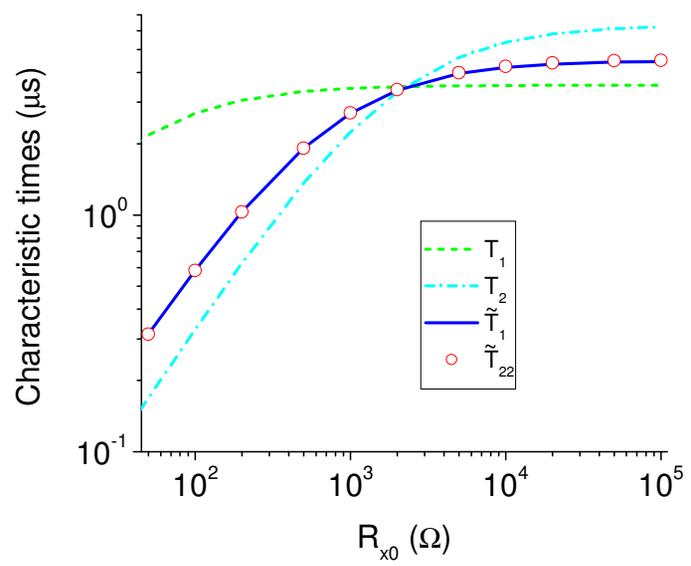
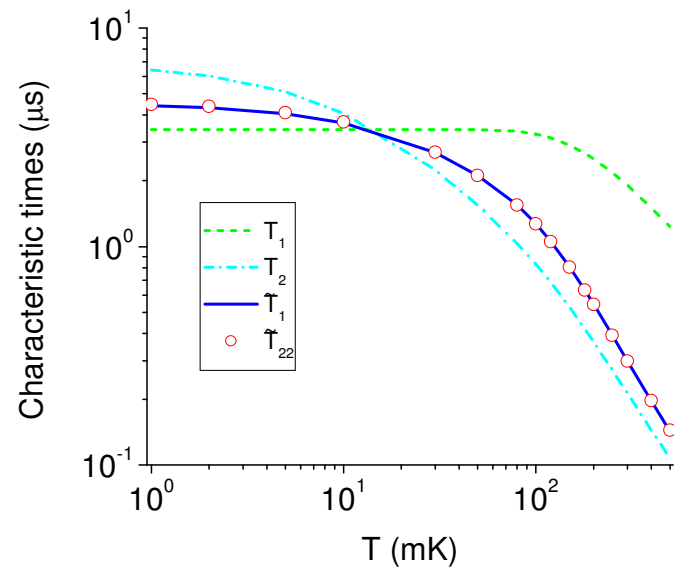
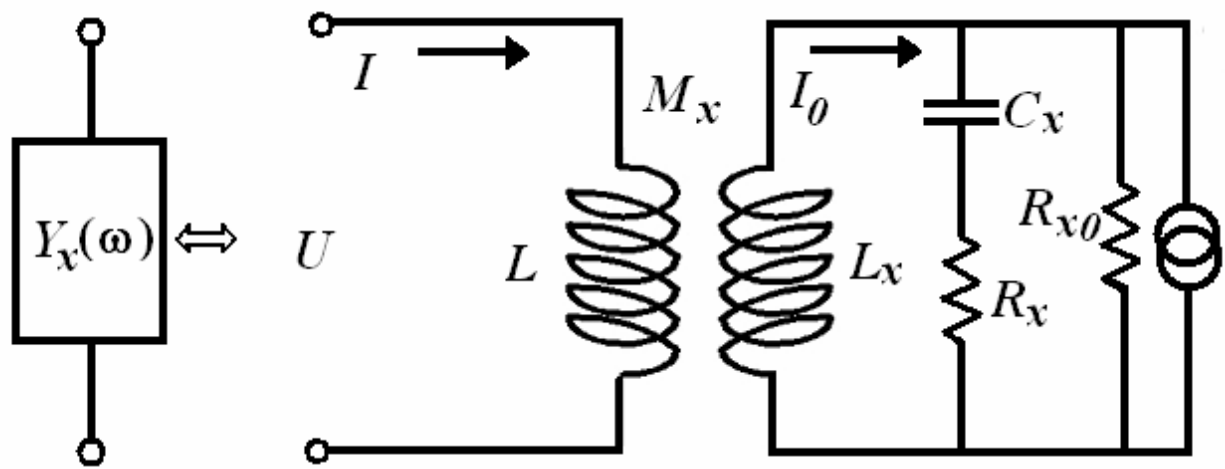
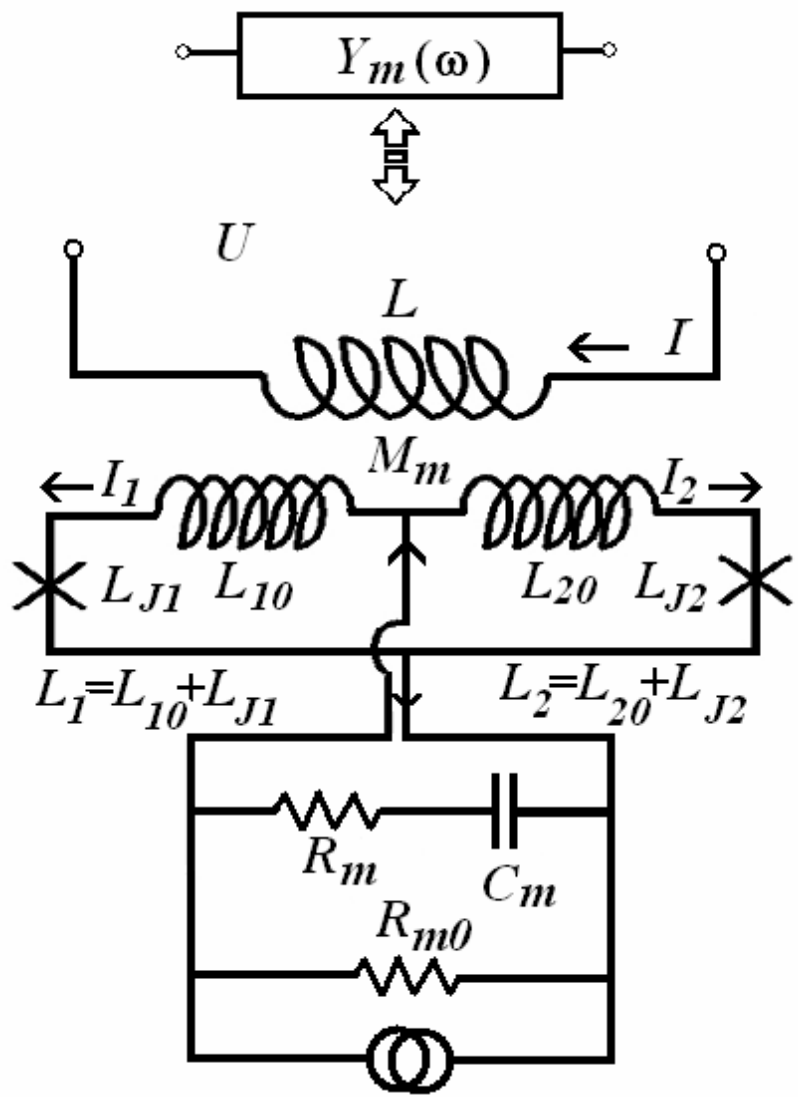
\end{document}